\documentclass[onecolumn,superscriptaddress,showpacs,preprintnumbers,amsmath,amssymb]{revtex4-1}

\usepackage{amsmath}
\usepackage{amssymb}
\usepackage{graphicx}
\usepackage{morefloats}
\usepackage{color}
\usepackage{blkarray}
\usepackage{verbatim}
\usepackage{hyperref}
\usepackage[centerlast,small]{caption}
\usepackage{subcaption}
\usepackage{xcolor}

\numberwithin{equation}{section}
\hypersetup{colorlinks=true, citecolor=orange, urlcolor=cyan, linkcolor=blue}

\begin{document}

\title{\bf{Information Recovery In Behavioral Networks}}
\author{Tiziano Squartini}
\affiliation{Istituto dei Sistemi Complessi, Universit\'a di Roma ``Sapienza'', P.le A. Moro 5, 00185 Rome (Italy)}
\author{Enrico Ser-Giacomi}
\affiliation{IFISC (CSIC-UIB), Instituto de F\'isica Interdisciplinar y Sistemas Complejos, Campus Universitat des les Illes Balears, E-07122 Palma de Mallorca (Spain)}
\author{Diego Garlaschelli}
\affiliation{Lorentz Institute for Theoretical Physics, University of Leiden, Niels Bohrweg 2, 9506 Leiden (Netherlands)}
\author{George Judge}
\affiliation{Graduate School and Giannini Foundation, University of California Berkeley, Berkeley, CA 94720 (United States)}
\date{\today}

\begin{abstract}
In the context of agent based modeling and network theory, we focus on the problem of recovering behavior-related choice information from origin-destination type data, a topic also known under the name of network tomography. As a basis for predicting agents' choices we emphasize the connection between adaptive intelligent behavior, causal entropy maximization and self-organized behavior in an open dynamic system. We cast this problem in the form of binary and weighted networks and suggest information theoretic entropy-driven methods to recover estimates of the unknown behavioral flow parameters. Our objective is to recover the unknown behavioral values across the ensemble analytically, without explicitly sampling the configuration space. In order to do so, we consider the Cressie-Read family of entropic functionals, enlarging the set of estimators commonly employed to make optimal use of the available information. More specifically, we explicitly work out two cases of particular interest: Shannon functional and the likelihood functional. We then employ them for the analysis of both univariate and bivariate data sets, comparing their accuracy in reproducing the observed trends.
\end{abstract}
\keywords{Binary Network; Adaptive Behavior; Causal Entropy Maximization; Inverse Problem; Information Theoretic Methods; Self-organized Systems}
\pacs{89.75.Da; 02.50.Le; 89.65.Ef}

\maketitle

\section{Introduction}

In this paper we focus on the problem of recovering behavior-related micro choice information from aggregate data. In particular, we consider origin-destination data, casting this problem as an inference problem concerning the prediction of flows on networks \cite{Vardi,Castro,Coates,Rubenstein}. We recognize that this type of data comes from dynamic, adaptive behavior systems involving interdependent micro components which give rise to an instantaneous, feedback-adaptive, world: as a result, such systems are non-deterministic in nature, involve information and uncertainty and are driven toward a certain, optimal, stationary state (see, for example, \cite{Annila,Georgescu-Roegen}). As a basis for predicting agents' choices, we cast this as a self-organized, equilibrium seeking system in the form of weighted and binary networks; we make use of information theoretic entropy-based methods to solve the ill-posed stochastic inverse problem and recover estimates of the unknown binary parameters.

\subsection{Binary Network Problem}

To go beyond traditional reductionist modeling and mathematical anomalies, we use a new paradigm that is developing under the name of Network Science (see, for example, \cite{Willinger,Barabasi} and the references contained therein). There are several things that make this approach attractive for information recovery in economics and in other social sciences: for example, in the economic-behavioral sciences everything seems to depend on everything else and this fits right into the interconnectedness simultaneity of the nonlinear (and dynamic) network paradigm. Another example is provided by microeconomic theory, where the network representation of markets arises quite naturally (in fact, in many ways markets and binary networks are equivalent - see \cite{Bargigli}). Finally, in terms of a methodology, network problems are consistent with the information theoretic approach to information recovery (see \cite{Mastrandrea2014a,Cimini2014b}). 

We seek an expression for the probabilities that the origin and the destination nodes are connected along a specific pathway in the statistical ensemble of possible pathways, without explicitly sampling the configuration space. Given information about the origin-destination network structure in the form of a matrix $\mathbf{A}$, the unknown pathway probabilities $p_{ij}$ must be estimated from aggregate flow data that may be noisy in nature. The number of unknown pathway parameters of the protocol matrix $\mathbf{A}$ is much larger than the number of measured aggregate origin-destination data points and, moreover, the components of the matrix $\mathbf{A}$ cannot be observed directly. This means that although the observed data are considered to be directly influenced by the values of model components, the observations only indirectly reflect the influence of the latter: as a result, the analyst must use indirect noisy observations to recover information on the unobserved vector of parameters. As a consequence, the relationship characterizing the effect of unobservable components on the observed data must be somehow inverted. This type of ill posed pure or stochastic inverse regularization problem cannot be solved by traditional econometric information recovery methods.

\subsection{Status Measure}

As we seek new ways to think about the causal adaptive behavior of complex and dynamic micro systems, we note that problems of this type may be re-formulated as problems of constrained entropy-maximization over the pathways. In other words, causal entropy maximization can be adopted as the systems status-measure and optimization criterion (following \cite{Wissner-Gross}). The result provides an exact expression for the occurrence of the unknown probabilities over the ensemble of pathways and yields the preferred probability distribution (see \cite{Presse}).

This permits us to recast a behavioral system in terms of path microstates where entropy reflects the number of ways a macrostate can evolve along a path of possible microstates: the more diverse the number of path microstates, the larger the causal path entropy. The result is a causal entropic force that captures self-organized equilibrium seeking behavior (see \cite{Wissner-Gross,Raine}). In other words, \emph{causal entropy maximization is a link that leads us to believe that a behavioral system with a large number of individuals, interacting locally and in finite time, is in fact optimizing itself}. We would like to stress that the optimization tendency characterizing behavioral systems is what qualifies entropy-based inference methods as the most correct ones to model such systems. The rationale beyond this lies in the nature of their adaptive behavior: agents tend to adapt behavior in line with an optimizing principle (as the maximization of the future, accessible paths diversity - also definable, more generally, as ``resources'' \cite{Wissner-Gross,Presse}), whence the need for a robust estimation procedure making the best use of the available information while disergarding any other arbitrary assumption. On the contrary, most behavioral economic-econometric models rest upon {\it ad hoc} assumptions which may lead to the identification and biased estimates of the unknown parameters, the underlying inference procedure and, in turn, the conclusions about the agents' behavior (see \cite{Bound,Angrist,DiPrete}).
\newline
\newline
\indent In the sections ahead we analyse systems within this framework, that permits the interpretation of adaptive economic behavior in terms of entropic functions: as a basis for solving micro-behavioral information recovery problems, we suggest an information theoretic family of entropic functions; to demonstrate applicability, we consider binary and weighted data sets and recover the optimum corresponding unknown probabilities. 

\section{Information Recovery Framework}

In developing a basis for the use of information theoretic (IT) methods to infer origin-destination networks flows, we focus on a stochastic ill posed inverse problem and the corresponding regularization method it implies (the pure, without-noise inverse problem is just a special case). In this context the Cressie-Read (CR) family of entropic functions \cite{Cressie,Read} provides a basis for linking the data and the unknown model parameters. 

This permits the researcher to exploit the statistical machinery of information theory to gain insights on the underlying adaptive behavior of a dynamic process from a system that may not be in equilibrium. This approach contrasts with the traditional approach to micro information recovery that rests on reductionist economic and econometric functional analysis and observational agent behavior data: however, precisely because of the nonlinear and ordinal nature of dynamic micro systems, the traditional approach is cumbersome in terms of identifying and expressing adaptive behavior.

We start introducing the CR multi parametric convex family of entropic functional measures \cite{Mittelhammer}: 

\begin{equation}
I(\mathbf{p}, \mathbf{q}, \gamma)=\frac{1}{\gamma(\gamma+1)}\sum_cp_c\left[\left(\frac{p_c}{q_c}\right)^\gamma-1\right].
\label{eq.1}
\end{equation}

In eq. \ref{eq.1}, $\gamma$ is a parameter that indexes members of the CR family, $p_c$'s represent the subject probabilities and the $q_c$'s are interpreted as reference (or prior) probabilities (the reason for indexing our coefficients with $c$ will be clarified in the following section). Being probabilities, the usual properties of $p_c,\:q_c\in [0,1],\:\forall\:c$, and $\sum_c p_c=1$, $\sum_c q_c=1$  are assumed to hold. As $\gamma$ varies the resulting CR estimator that minimize the divergence between $\mathbf{p}$ and $\mathbf{q}$ exhibits a qualitatively different behavior that includes, as noteworthy examples, the Kullback-Leibler measure (in the limit as $\gamma\rightarrow0$ as Shannon entropy and in the limit as $\gamma\rightarrow-1$ as the likelihood functional) and, in a binary context, the logistic distribution-divergence (see \cite{Gorban2003}). 

In other words, the CR family of power divergences is a class of additive convex functions that encompasses a broad family of test statistics, in turn representing a broad family of functional relationships within a moments-based estimation context. In addition, the CR measure exhibits proper convexity in $\mathbf{p}$, for all values of $\gamma$ and $\mathbf{q}$, and embodies characteristics such as additivity and invariance with respect to a monotonic transformation of the divergence measures. In the context of extremum metrics, the CR family represents a flexible family of pseudo-distance measures from which to derive empirical probabilities.

\section{Integer Versions of the CR Family}

%In what follows we consider the three values $\gamma=-1,0,1$, corresponding respectively to the \emph{likelihood functional}, the \emph{Shannon functional} and the \emph{Euclidean functional}. In the limit as $\gamma\rightarrow 0$

In what follows we consider the two values $\gamma=-1,0$, corresponding respectively to the \emph{likelihood functional} and the \emph{Shannon functional}. In the limit as $\gamma\rightarrow 0$

\begin{equation}
\lim_{\gamma\rightarrow0}I(\mathbf{p}, \mathbf{q}, \gamma)=\sum_{c}p_c\ln\left(\frac{p_c}{q_c}\right)
\label{shaeq}
\end{equation}

\noindent the Kullback-Leibler divergence between $\mathbf{p}$ and $\mathbf{q}$ is obtained. The particular case of a uniform prior, $q_c=1/C$, allows us to recover the usual form of (minus) the \emph{Shannon entropy} of the $\mathbf{p}$ distribution: $I\left(\mathbf{p}, \frac{1}{C}, 0\right)=\sum_cp_c\ln p_c+\ln C$. In the limit as $\gamma\rightarrow -1$ provides the second functional of our list

\begin{equation}
\lim_{\gamma\rightarrow -1}I(\mathbf{p}, \mathbf{q}, \gamma)=\sum_cq_c\ln\left(\frac{q_c}{p_c}\right)
\end{equation}

\noindent the Kullback-Leibler divergence between $\mathbf{q}$ and $\mathbf{p}$. The particular case of uniform prior $q_c=1/C$ allows us to recover the usual form of (minus) the \emph{likelihood function} of the $\mathbf{p}$ distribution: $I\left(\mathbf{p}, \frac{1}{C}, -1\right)=-\sum_{c}\frac{\ln p_c}{C}-\ln C$. %The value $\gamma=1$ provides the third functional

%\begin{equation}
%I(\mathbf{p}, \mathbf{q}, 1)=\sum_{c}\frac{p_c}{2}\left(\frac{p_c}{q_c}-1\right)
%\end{equation}

%\noindent known as \emph{Euclidean functional} which, in the particular case of uniform prior $q_c=1/C$, allows us to recover the form $I\left(\mathbf{p}, \frac{1}{C}, 1\right)=\frac{C}{2}\sum_{c}p_c\left(p_c-\frac{1}{C}\right)$.

We stress that while the Shannon functional has been already employed for the analysis of univariate and bivariate data sets, the likelihood functional case has not been explicitly worked out yet, thus representing the major contribution of this paper to the analysis of behavioral networks.

\begin{figure}[h!]
\begin{center}
\includegraphics[width=0.35\textwidth]{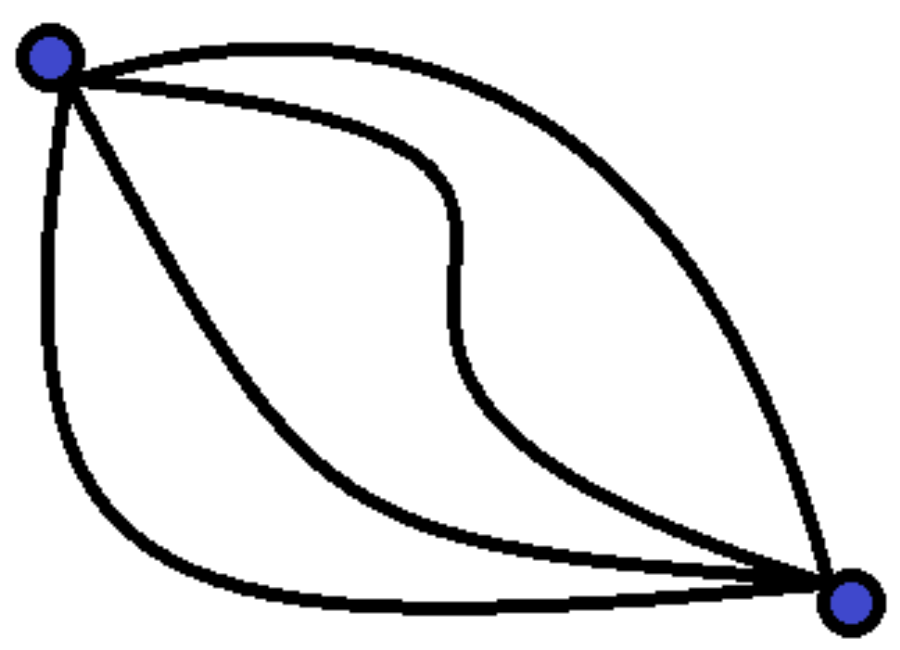}
\caption{A schematic representation of an origin-destination network. Blue dots represent the origin and the destination nodes. Connections between them represent the ensemble of pathways described by the probability distribution $\{p_c\}_{c=1}^C$. The CR family allows one to determine the probability coefficients $p_c,\:\forall\:c$ by making use of the available partial information, i.e. aggregate data on traffic volumes.}
\label{fig1}
\end{center}
\end{figure}

\section{Network Behavior Recovery}

To demonstrate the applicability of our approach in the binary network area, an example may be useful. Consider the problem of determining least-time, point-to-point traffic flows between sub-networks, when only aggregate origin-destinations volumes are known (see fig. \ref{fig1}). In many ways this is like a transportation network, with the emphasis on design and efficiency in routing the traffic flows (see \cite{Castro} and the references therein), exactly as in an economic-behavioral network the efficiency of information flow is predicated on discovering, or designing, protocols that efficiently route information. The research question concerns the prediction of the volume of flows on the pathways, given a set of measures taken along them. 

If we indicate by $\mathbf{y}$ the $R$-dimensional vector of observed fluxes and by $\mathbf{x}$ the $C$-dimensional vector of intermediate measures, the ``activity'' of an origin-destination network can be summed up by writing 

\begin{equation}
\mathbf{y}=\mathbf{A}\mathbf{x}
\label{eq.1}
\end{equation}

\noindent where $\mathbf{A}$ is an $R\times C$  rectangular matrix, encoding the information about connections. Thus, our problem translates into estimating $\mathbf{x}$ on the basis of the $R$, available components of $\mathbf{y}$ and the connection structure $\mathbf{A}$. The ill-posed nature of the problem is such that the inversion of eq. \ref{eq.1} is not feasible: the number of unknowns is greater than the number of known data, i.e. $R<C$. In this case, one can resort to the information theoretic methodology for solving problems of inference on the basis of partial information (see \cite{Cho2014,Judge2012a,Ziebart2010,Ziebart2013}). In order to implement, the problem unknowns have to be interpretable as probabilities and estimated on the basis of some known distribution moments. In our case, this can be easily achieved by dividing both sides of \ref{eq.1} by $x_{tot}\equiv\sum_cx_c$:

\begin{equation}
\frac{\mathbf{y}}{x_{tot}}\equiv\mathbf{r}=\mathbf{A}\mathbf{p}\equiv\mathbf{A}\frac{\mathbf{x}}{x_{tot}}
\label{eq.2}
\end{equation}

\noindent where $\mathbf{y}$ and $\mathbf{A}$ are known, $\mathbf{p}$ is unknown and $\sum_c p_c=1$. We have thus rewritten \ref{eq.1} in terms of \emph{fractions of fluxes} distributed across the $C$ channels and interpret them as unknown probabilities. Notice that this peculiar definition of probability coefficients induces a distribution on the set of pathways, that play the role of an \emph{ensemble} and allows us to restate the problem of predicting the fluxes on origin-destination networks as a (more) general problem of statistical inference. We can now may make use of the CR family of entropic divergence measures and write the problem as the following constrained optimization problem:

\begin{equation}
\mathcal{L}\equiv I(\mathbf{p},\mathbf{q},\gamma)-\theta_0\left[\sum_c p_c-1\right]-\sum_\alpha\theta_\alpha\left[\sum_c p_cA_{\alpha c}-r_\alpha\right]
\label{eq.3a}
\end{equation}

In particular, since the functional $I$ is a divergence, the Lagrangean function has to be minimized with respect to the vector of coefficients $\mathbf{p}$. This gives us the desired coefficients $\{p_c\}_{c=1}^C$ as functions of the Lagrangean multipliers, $p_c=p_c(\vec{\theta}),\:\forall\:c$. Once found, the parametric probability coefficients must be substituted back into $\mathcal{L}$, in order to obtain a quantity which is a function of the unknowns solely: $\mathcal{L}(\vec{\theta})$. The last step of our procedure prescribes the optimization of the function $\mathcal{L}(\vec{\theta})$ (see the Appendix, ``Univariate data sets'' section, for the detailed calculations).

A similar problem is faced whenever a whole matrix of probability coefficients (and not a simple vector), $\mathbf{P}$, is considered. Problems of this type can be formulated in much the same way, by writing the equation

\begin{equation}
\mathbf{y}'=\mathbf{x}'\mathbf{P}
\end{equation}

\noindent thus mimicking \ref{eq.1}. As we will  show, treating $\mathbf{y}'$ and $\mathbf{x}'$ as known vectors allows us to succesfully also tackle this second type of problem (see the Appendix, ``Bivariate data sets'' section, for the detailed calculations).
\newline
\newline
\indent These are just the solutions to a standard problem when a function must be inferred from insufficient sample-data information. Thus network inference and monitoring problems have a strong resemblance to an inverse problem in which key aspects of a system are not directly observable (for details on the use of information theoretic entropic methods for this type of network information flow problems see also \cite{Cho2014,Cho2006,Ziebart2010,Ziebart2013}).

\section{Applications}

To test the effectiveness of our method, in what follows we analyze two aggregate data sets (for which origin-destination traffic volumes were collected), the first one concerning traffic on a local area network and the second one concerning consumers' choices of complementary products.

\begin{figure}[t!]
\centering
\includegraphics[width=0.5\textwidth]{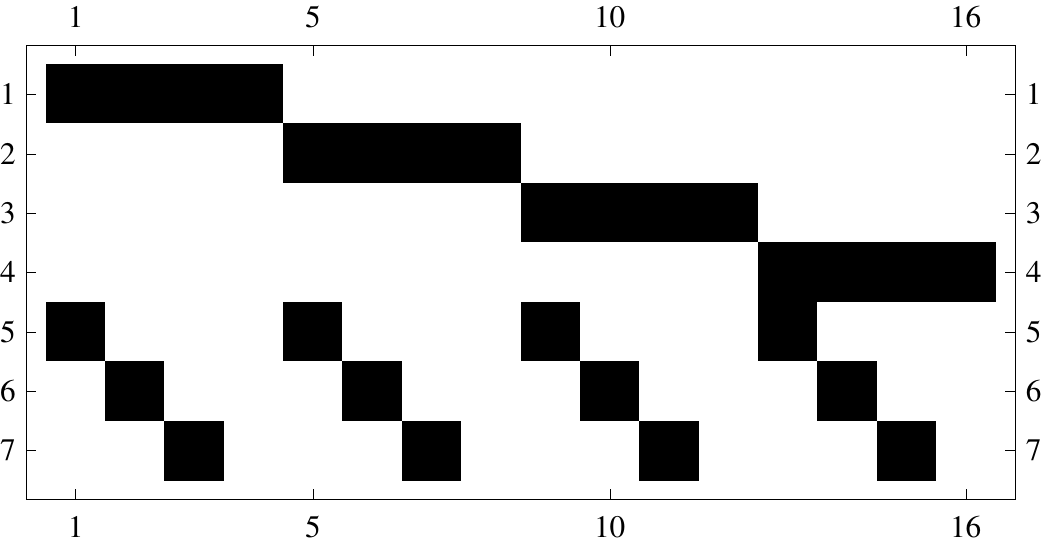}
\caption{Pictorial matrix representation of a local area network at Bell Labs (black squares represent ones, white squares represent zeros - see \cite{Cho2014,Airoldi}), composed by four subnetworks (fddi, corp, local and switch) communicating via a router. The network topology we consider yields 7 observed aggregate traffic volumes and 16 origin-destination traffic volumes.}
\label{fig2a}
\end{figure}

\paragraph{Bell Labs data.} The first data set involves traffic volumes on a local area network at Bell Labs (see \cite{Cho2014,Airoldi}) whose routing matrix is reported in fig. \ref{fig2a}. The network topology we consider here yields 7 observed aggregate traffic volumes and 16 origin-destination traffic volumes to be estimated. Aggregate volumes were measured every five minutes, over one day, on the Bell Labs corporate network, resulting in a set of measurements of 287 time points.

\paragraph{Complementary products.} The second data set comes from an economic case-study and relates to consumers' behavior in the purchase of eggs and bacon (see \cite{Cho2014,Crackel}). In particular, data consist of a sample of 548 independent households and the purchased products at the market, recorded over 4 consecutive trips. For each trip, it was recorded whether or not the household purchased eggs, bacon or both: the matrix entries represent the number of times a given customer purchased bacon and eggs over the course of the 4 trips (as reported in table \ref{table1} - see also \cite{Crackel}).

\begin{table}[h!]
\centering
\caption{Observed bivariate distribution of the number of times bacon and eggs were purchased on four consecutive shopping trips (see \cite{Cho2014,Crackel}).}
\begin{tabular}{c|ccccc|c}
\noalign{\smallskip}
$$ & $$ & $$ & $\mbox{Eggs}$ & $$ & $$ & $$\\
$\mbox{Bacon}$ & $0$ & $1$ & $2$ & $3$ & $4$ & $\mbox{Total}$\\
\hline
$0$ & $254$ & $115$ & $42$ & $13$ & $6$ & $430$\\

$1$ &$34$ & $29$ & $16$ & $6$ & $1$ & $86$\\

$2$ &$8$ & $8$ & $3$ & $3$ & $1$ & $23$\\

$3$ &$0$ & $0$ & $4$ & $1$ & $1$ & $6$\\

$4$ &$1$ & $1$ & $1$ & $0$ & $0$ & $3$\\
\hline
$\mbox{Total}$ &$297$ & $153$ & $66$ & $23$ & $9$ & $548$
\end{tabular}
\label{table1}
\end{table}

\begin{figure}[t!]
\centering
\includegraphics[width=0.48\textwidth]{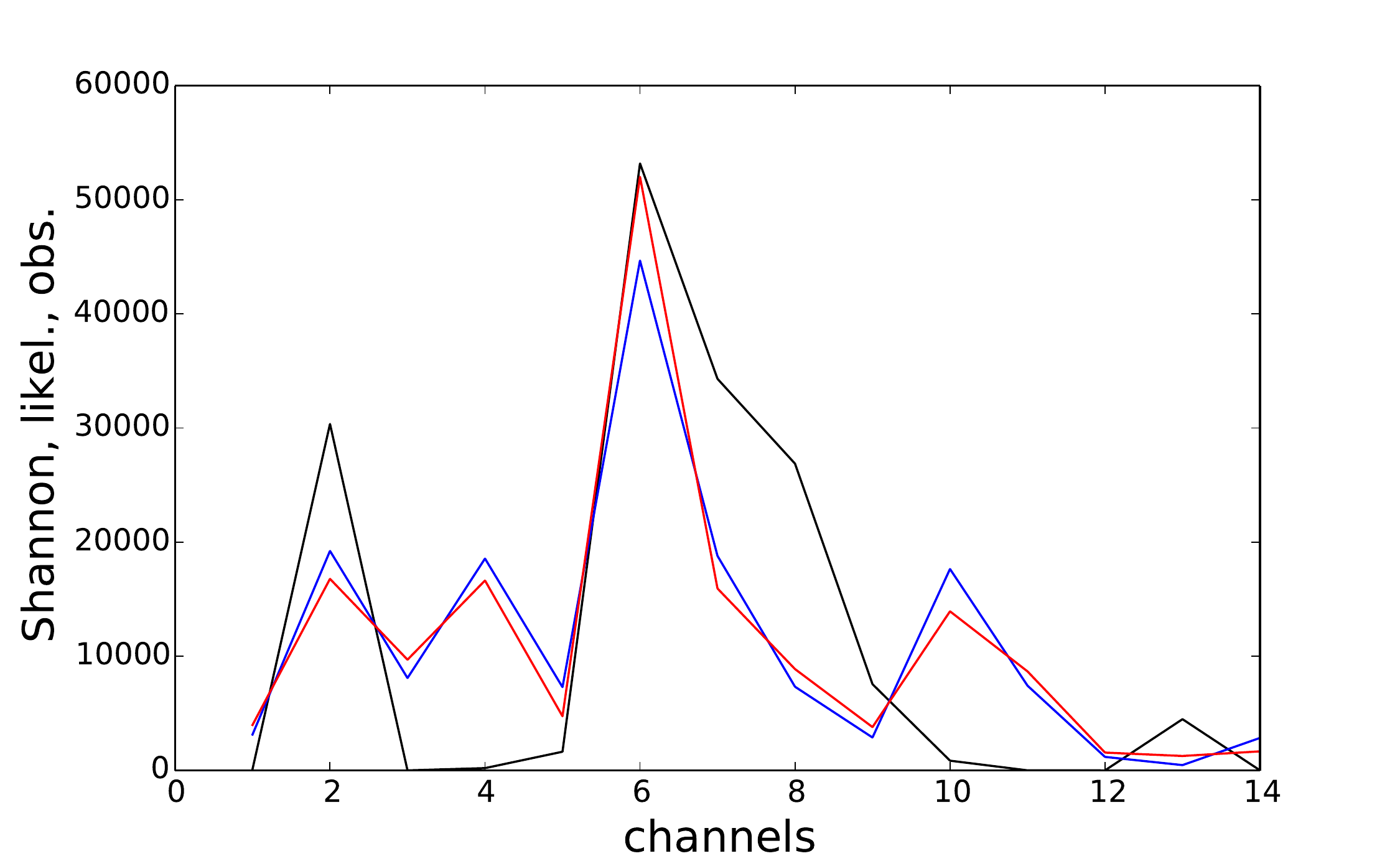}
\includegraphics[width=0.48\textwidth]{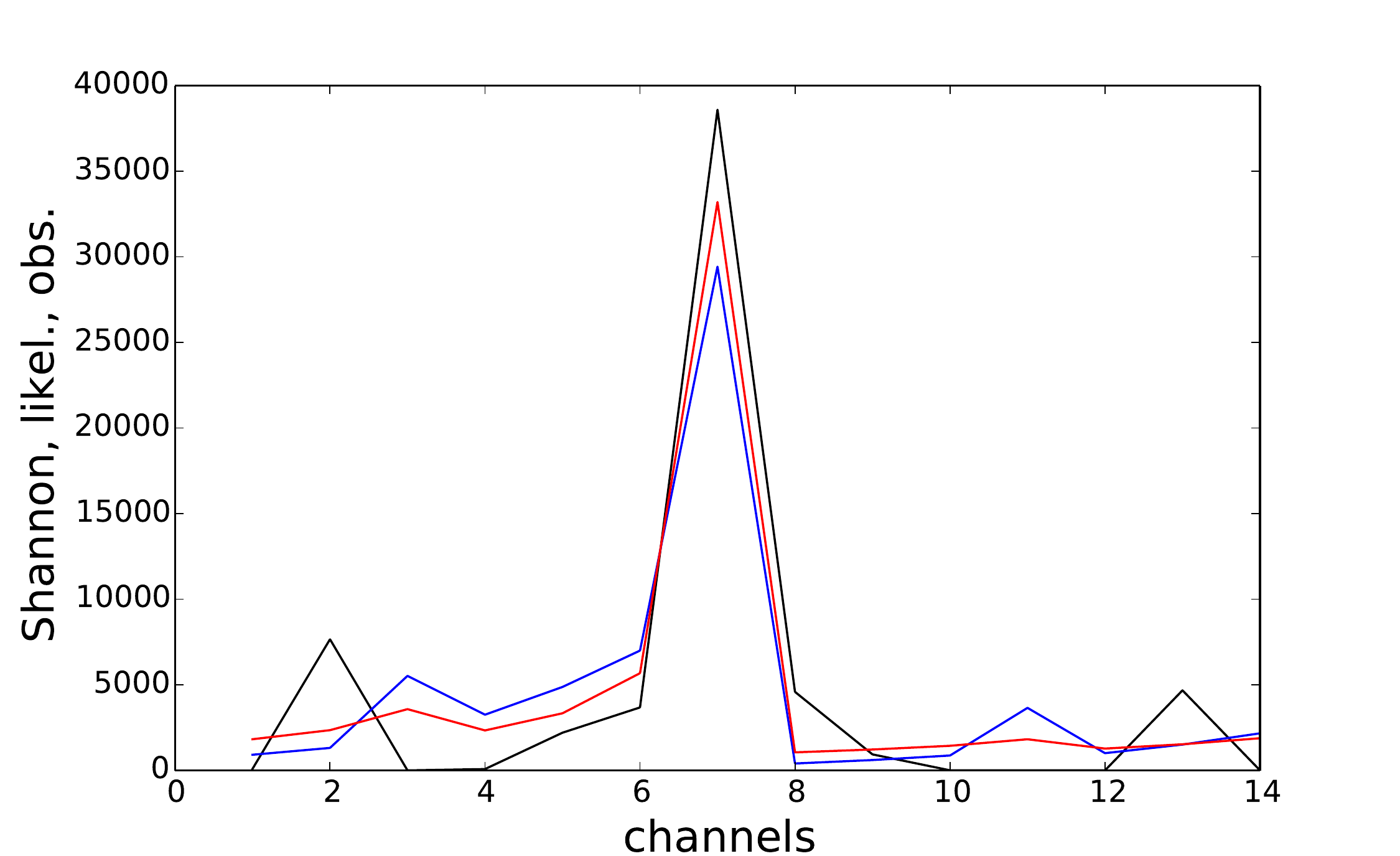}
\includegraphics[width=0.48\textwidth]{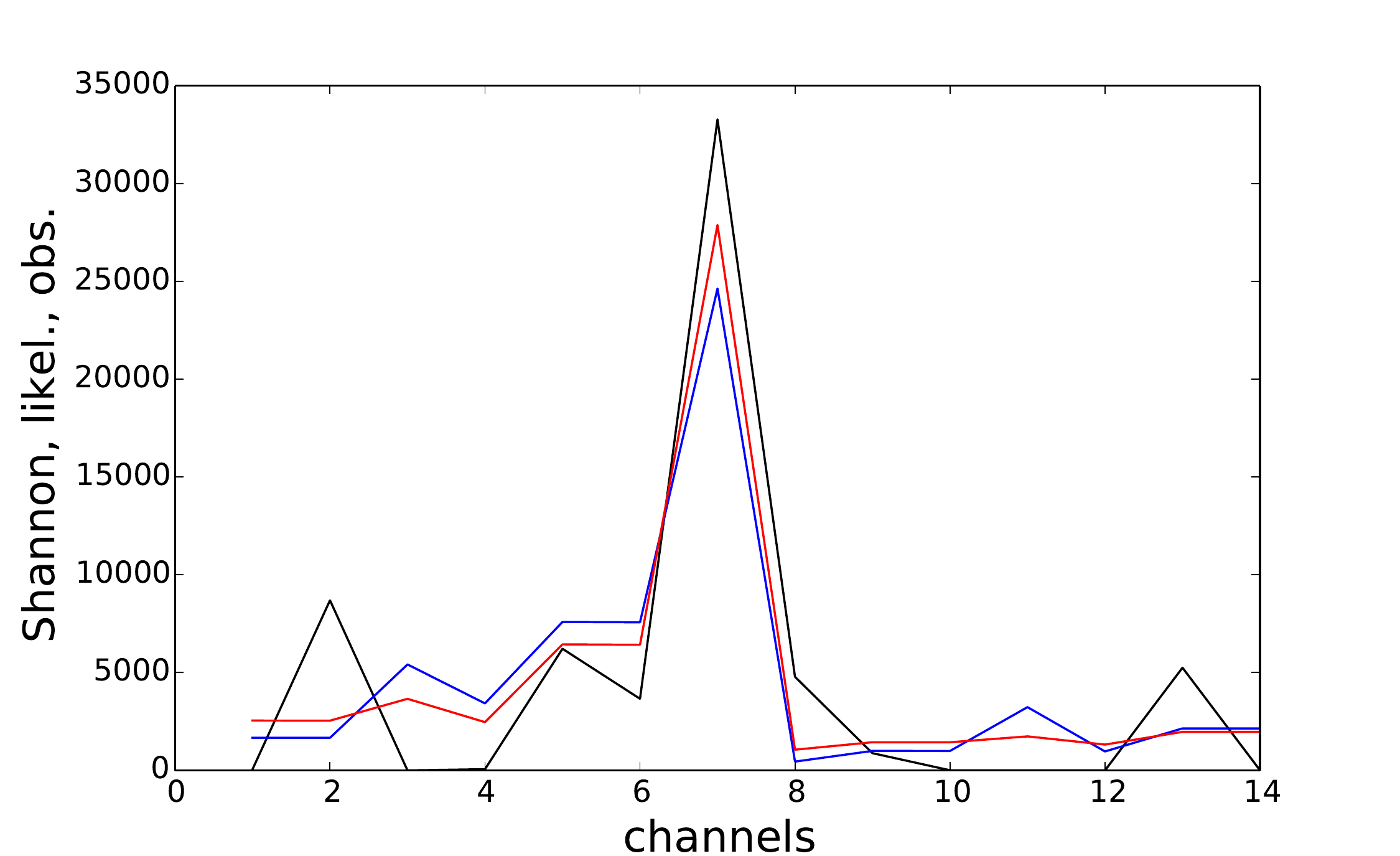}
\includegraphics[width=0.48\textwidth]{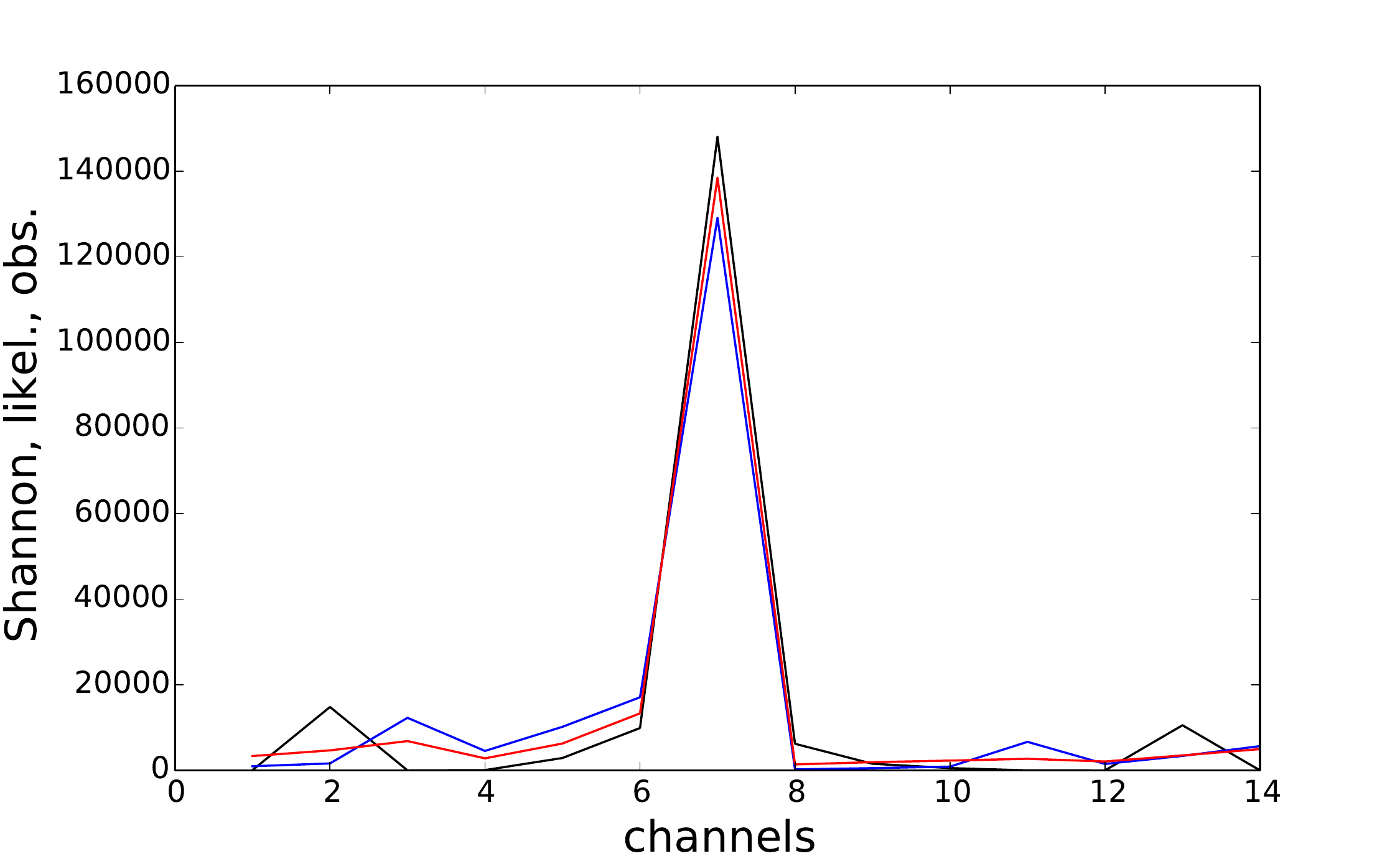}
\includegraphics[width=0.48\textwidth]{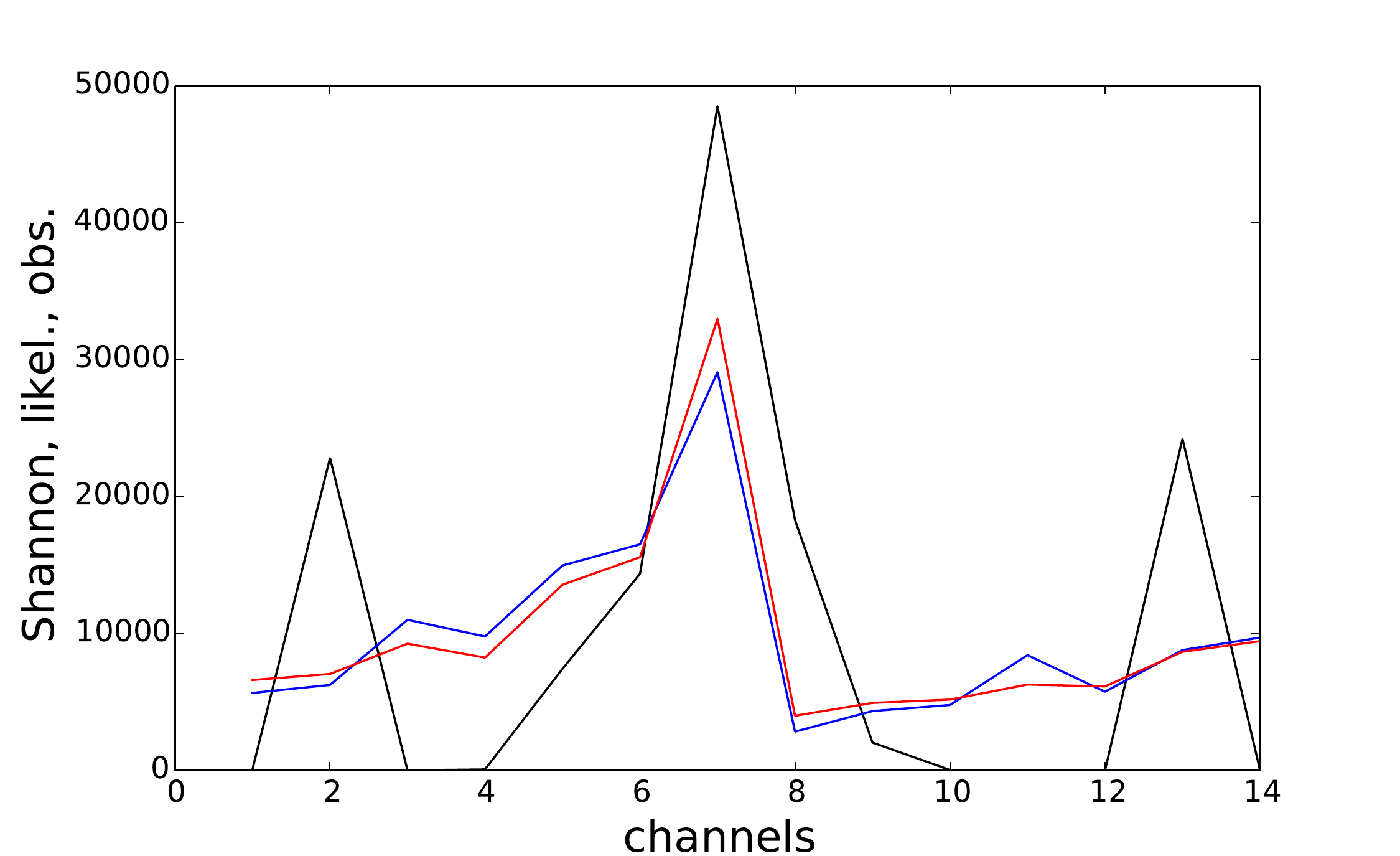}
\includegraphics[width=0.48\textwidth]{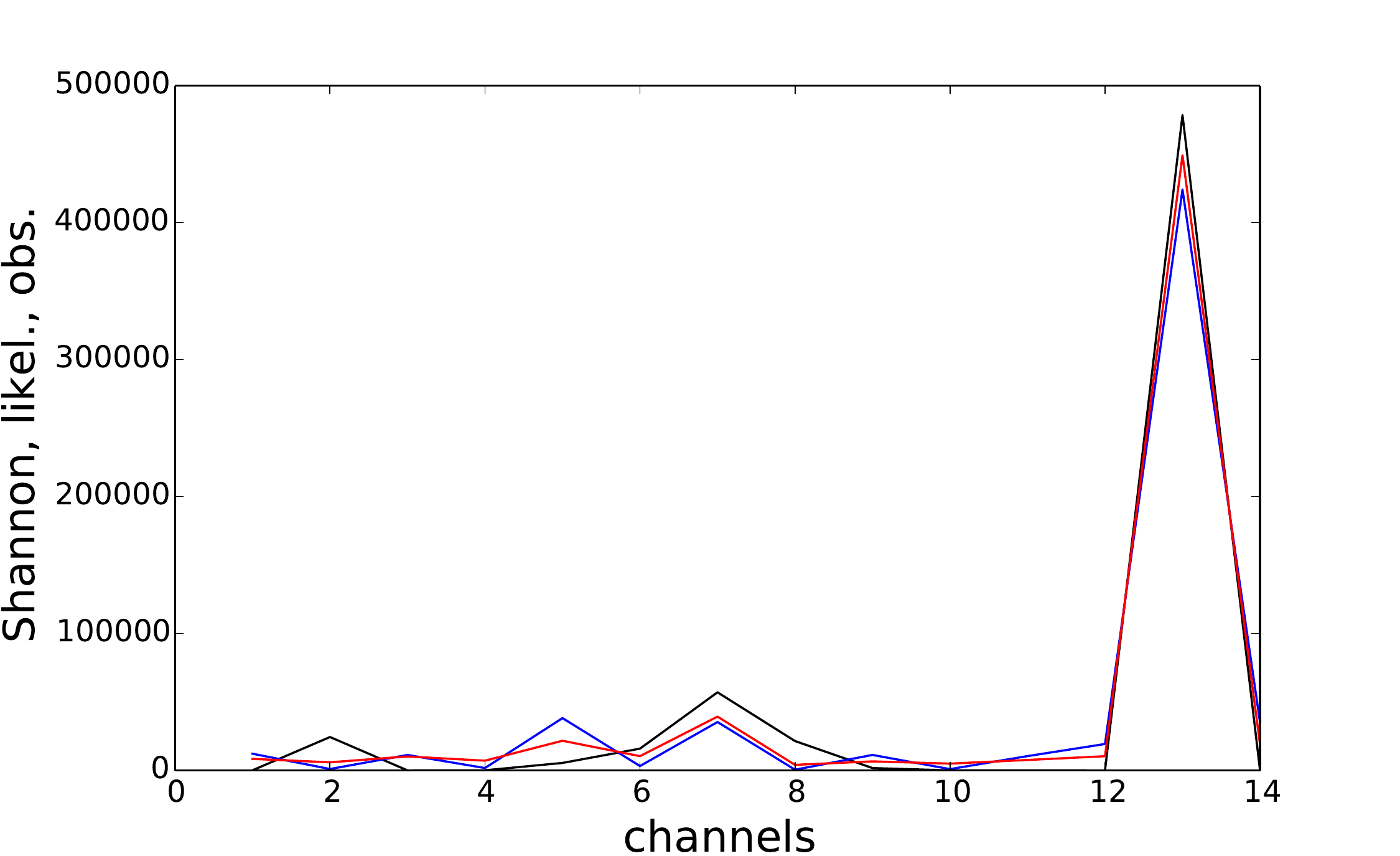}
\includegraphics[width=0.48\textwidth]{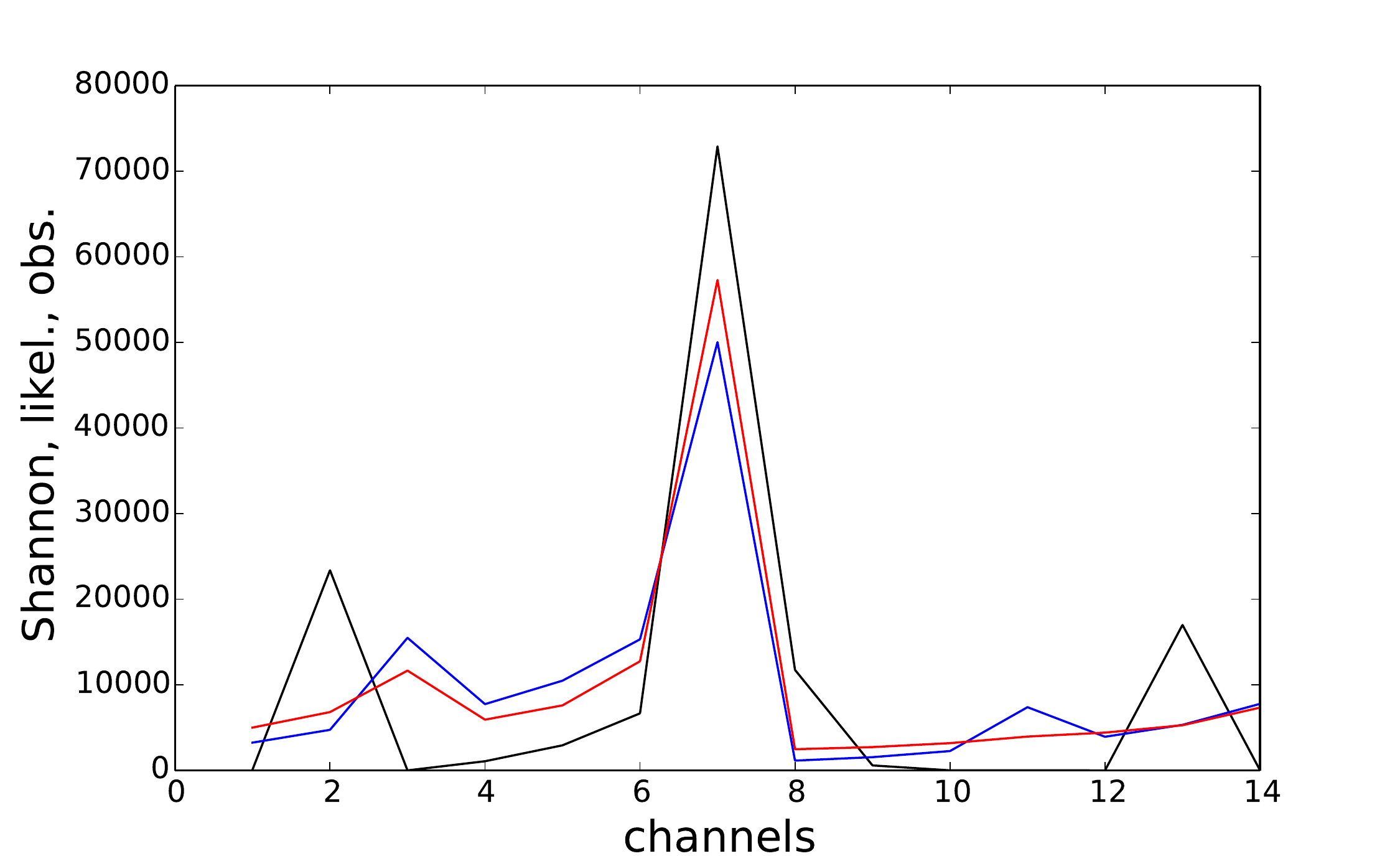}
\includegraphics[width=0.48\textwidth]{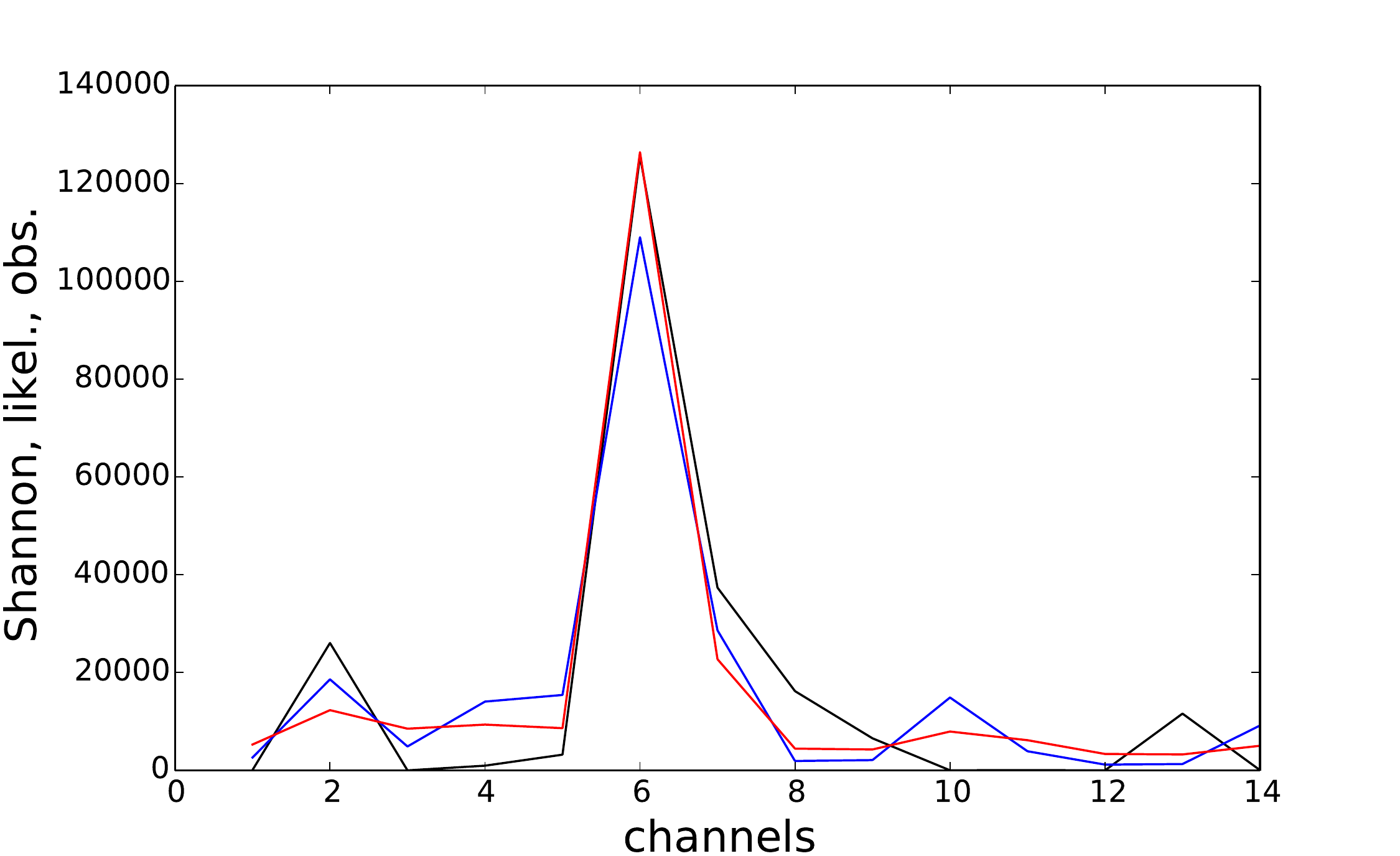}
\caption{Analysis of Bell Labs data corresponding to ten chosen time points. The number of the channel is reported on the x-axis. Observed and estimated $\mathbf{x}$ are reported on the y-axis. Colors referes to: observed data (black trend), our estimation based on Shannon functional (blue trend), our estimation based on the likelihood functional (red trend).}
\label{fig4}
\end{figure}

\subsection{Bell Labs data}

\begin{figure}
\includegraphics[width=0.508\textwidth]{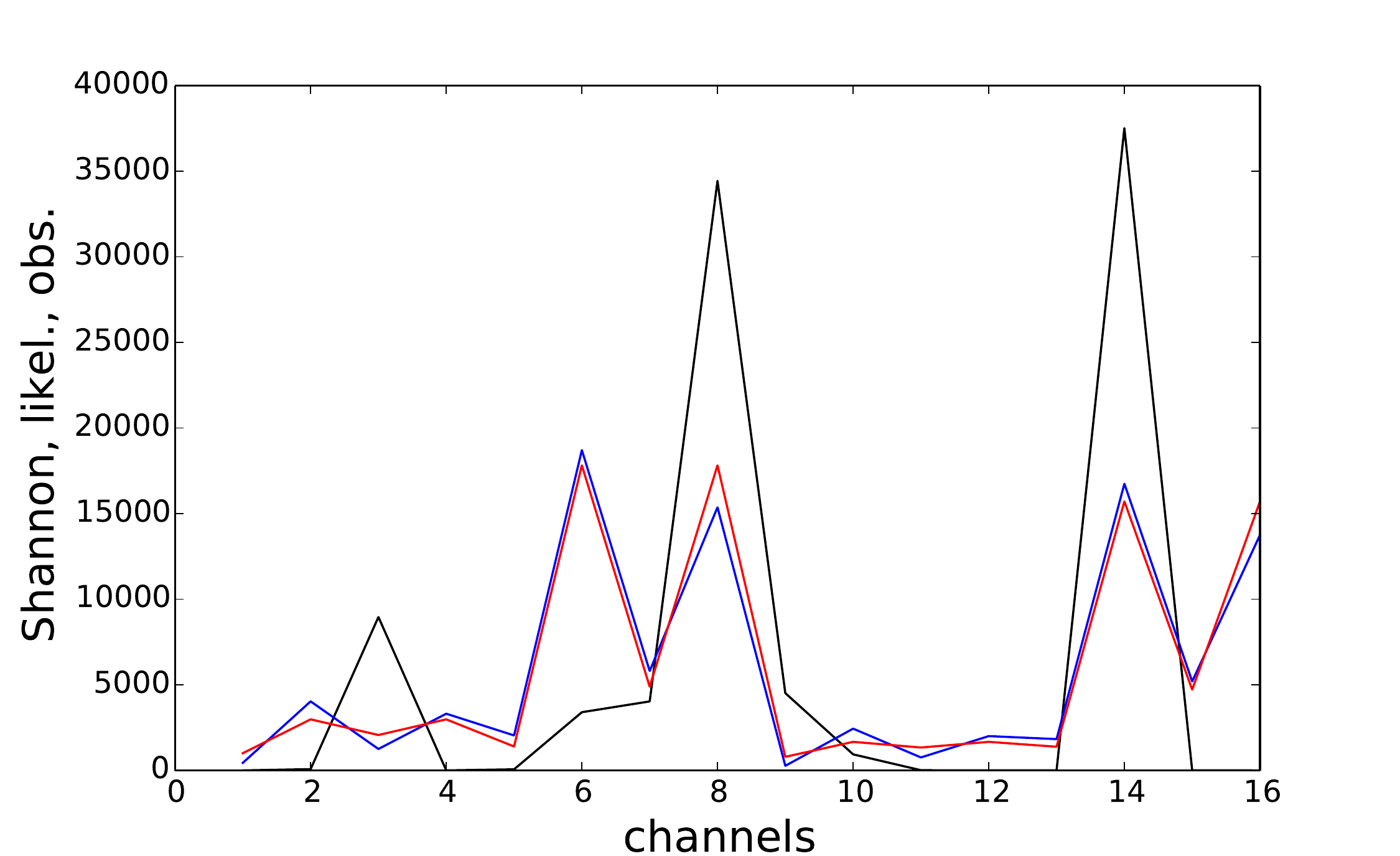}
\includegraphics[width=0.48\textwidth]{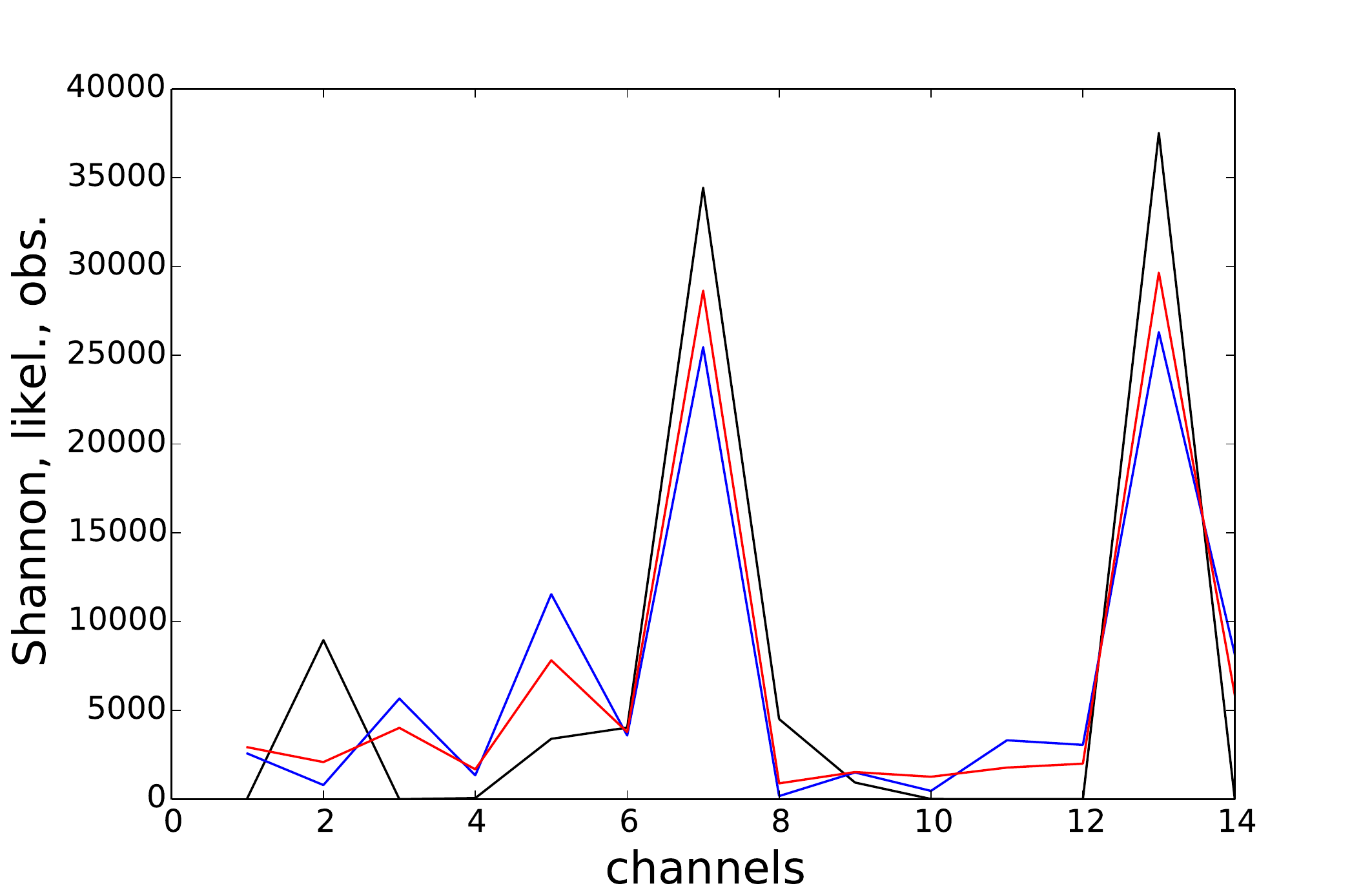}
\caption{Analysis of Bell Labs data for the $90$th time point. The number of the channel is reported on the x-axis. Observed and estimated $\mathbf{x}$ are reported on the y-axis. Colors referes to: observed data (black trend), our estimation based on Shannon functional (blue trend), our estimation based on the likelihood functional (red trend). Left panel: zero traffic flows are included in the data set. Right panel: zero traffic flows are excluded from the data set.}
\label{fig5}
\end{figure}

The analysis of Bell Labs data is illustrated in figs. \ref{fig4}, \ref{fig5}. The panels report what we have called ``channel plots'', showing the label of each origin-destination pattern (or channel) on the x-axis and the traffic volumes measured and estimated on it, on the y-axis. Black trends represent the observed traffic volumes and colored trends represent the expected traffic volumes, predicted via our procedure: blue trends represent the predictions obtained by using Shannon functional, red trends represent the predictions obtained with the empirical likelihood functional. Each panel corresponds to a given time point, chosen among the 287 available possibilities.

As a general comment, the predictions of both functionals reproduce the majority of the observed trends satisfactorily, with the likelihood functional performing slightly better than Shannon functional whose estimates, in some cases, show larger discrepancies. Moreover, the performance of both functionals improves when single peaks are registered on a single channel, accompanied by small traffic volumes on the others. However, at night, whenever the latter are exactly zero the agreement between our estimates and observations seems to deteriorate: as shown in the left panel of fig. \ref{fig5}, if zero traffic flows happen to be measured on some line, both Shannon and the likelihood functionals predict smaller peaks and larger values for the neighboring lines. 

A solution to improve the predictions accuracy is to explicitly exclude zero values from our dataset. This can be achieved by considering a reduced $\mathbf{x}$ vector and a reduced $\mathbf{A}$ matrix without the 1st and the 16th columns, i.e. precisely those contributing to the values $x_1=x_{16}=0$. The right panel of fig. \ref{fig5} shows how much the accuracy of our method is improved: notice how peaks are reproduced much better now and traffic values on the neighboring lines are predicted to be much smaller than the former, as observed values confirm. The predicted trends in fig. \ref{fig4} are calculated by adopting the same criterion, i.e. explicitly excluding the zero values on the extreme channels.

%The Euclidean functional (not shown in figs. \ref{fig4}, \ref{fig5}) performs worst: its functional form often leads to problems of numerical convergence and, in any case, is characterized by the largest discrepancies between expected and observed patterns. 

\subsection{Complementary products}

The result of the application of our information recovery method to the ``eggs and bacon'' data set is shown in table \ref{table3}. Since the anaysis concerns a bivariate network, the predictions of our functionals concern the matrix entries, estimated from the available rows and columns totals (see the Appendix, ``Bivariate data sets'' section, for the detailed calculations).

Table \ref{table3} depicts the predictions based on Shannon functional, the likelihood functional and the Euclidean functional. In order to further condense the information, we have also calculated the correlation coefficient between each observed row and the corresponding expected one, reporting the obtained values in the last entry of each row of table \ref{table3}. The correlation coefficients are high for all the three functionals, which predict close values to the observed ones. 

A closer inspection of table \ref{table3} reveals that, as for the Bell Labs data set, the rows with the zeros are still the most problematic ones. However, the likelihood functional performs better than Shannon one: the predicted entries are closer to the real ones and the correlation coefficients are higher.

\begin{table*}[b!]
\centering
\caption{Expected bivariate distribution of the number of times bacon and eggs were purchased on four consecutive shopping trips (see \cite{Cho2014,Crackel}).}
\begin{tabular}{|c|ccccc|c|}
\hline
$$ & $$ & $$ & $\mbox{Shannon functional}$ & $$ & $$ & $$\\
\hline
$$ & $$ & $$ & $\mbox{Eggs}$ & $$ & $$ & $$\\
$\mbox{Bacon}$ & $0$ & $1$ & $2$ & $3$ & $4$ & $r$\\
\hline
$0$ & $262.378$ & $122.478$ & $40.468$ & $4.65702$ & $0.0191661$ & $0.999453$\\

$1$ &$27.3702$ & $23.502$ & $18.8328$ & $12.2212$ & $4.0738$ & $0.970398$\\

$2$ &$5.38417$ & $5.16918$ & $4.87188$ & $4.33981$ & $3.23497$ & $0.86233$\\

$3$ &$1.25404$ & $1.24078$ & $1.22175$ & $1.18545$ & $1.09798$ & $-0.0718339$\\

$4$ &$0.613532$ & $0.61028$ & $0.605583$ & $0.596516$ & $0.574089$ & $0.847078$\\
\hline
\hline
$$ & $$ & $$ & $\mbox{Likelihood functional}$ & $$ & $$ & $$\\
\hline
$$ & $$ & $$ & $\mbox{Eggs}$ & $$ & $$ & $$\\
$\mbox{Bacon}$ & $0$ & $1$ & $2$ & $3$ & $4$ & $r$\\
\hline
$0$ & $258.603$ & $118.489$ & $40.1875$ & $9.81096$ & $2.90897$ & $0.99991$\\

$1$ &$30.4192$ & $26.7046$ & $18.5562$ & $7.63744$ & $2.68261$ & $0.993516$\\

$2$ &$6.02486$ & $5.86333$ & $5.34772$ & $3.78732$ & $1.97677$ & $0.850168$\\

$3$ &$1.32087$ & $1.31294$ & $1.28519$ & $1.1694$ & $0.911598$ & $0.019223$\\

$4$ &$0.631723$ & $0.629903$ & $0.623446$ & $0.594872$ & $0.520056$ & $0.824691$\\
\hline
%\hline
%$$ & $$ & $$ & $\mbox{Euclidean functional}$ & $$ & $$ & $$\\
%\hline
%$$ & $$ & $$ & $\mbox{Eggs}$ & $$ & $$ & $$\\
%$\mbox{Bacon}$ & $0$ & $1$ & $2$ & $3$ & $4$ & $r$\\
%\hline
%$0$ & $265.873$ & $127.823$ & $44.4184$ & $3.19533$ & $14.916$ & $0.997586$\\

%$1$ &$24.1851$ & $18.6631$ & $15.3269$ & $13.678$ & $14.1468$ & $0.908502$\\

%$2$ &$5.09961$ & $4.70465$ & $4.46603$ & $4.34809$ & $4.38162$ & $0.863061$\\

%$3$ &$1.234$ & $1.20712$ & $1.19088$ & $1.18286$ & $1.18514$ & $-0.490478$\\

%$4$ &$0.6085$ & $0.60178$ & $0.597721$ & $0.595714$ & $0.596285$ & $0.687752$\\
%\hline
\end{tabular}
\label{table3}
\end{table*}

\section{Some summary comments}

This paper represents a contribution to the study of behavioral information recovery for self-organizing systems. The approach we proposed questions the use of traditional information recovery methods (see \cite{Presse}), stressing the connections between adaptive behavior and causal entropy maximization (see \cite{Wissner-Gross}) in self organizing systems. This intuition can be formalized by implementing the procedure we propose, resting on the optimization of a class of entropic functionals under the constraints provided by the available information. Remarkably, other studies have presented results compatible with this view, i.e. that the real word is well approximated by maximum entropy ensembles where only partial information is used to reconstruct the entire system (see \cite{Mastrandrea2014a,Cimini2014b,Squartini2014}). 

The class of entropic functionals employed in this work is known as Cressie-Read family, which not only constitutes the analytical basis of our analysis but also represents a solution to the issue of solving ill-posed inverse problems by formally treating them as inference problems. Our results indicate that the performance of functionals constituting the CR family may vary significantly: in some cases, the likelihood functional (to the best of our knowledge, explicitly worked out here for the first time) provides the best performance; in others, it is outperformed by the Shannon functional. This indicates these two functionals are the ones making the best possible use of the available information, predicting the closest values to the observed ones. 

In order to suggest applicability of our procedure, we have considered behavioral problems within the framework of network theory. The results we obtained not only indicate the effectiveness of our algorithm (applicable to univariate as well as bivariate data sets and for both \emph{reproducing available data} and \emph{predicting unavailable data}), but also demonstrate that networks are a useful way to present micro behavioral systems. In this context, the perspective proposed by our study can be enlarged by considering each node as a network on its own, a possibility which would simplify the task of modelling evolving networks, such as in the case of a growing economy, where a larger number of (adapting) nodes appear.

Given the importance of recovering dynamic economic behavioral information, a natural question arises about the continued use of traditional regularization information recovery methods as a solution basis for traditional pure and stochastic inverse type problems. For this reason, the next step is to extend the concept of adaptive-optimizing behavior and apply it (within the information theoretic framework) in the context of a range of micro economic settings, thus opening the promising perspective of turning the descriptive character of behavioral disciplines into a more quantitative one.

\section{Appendix}

\subsection{Univariate data sets}

As previously noted, eq. \ref{eq.2} induces a distribution on the ensemble of pathways. In other words, eq. \ref{eq.2} allows us to restate the problem of predicting the fluxes on origin-destination networks as a (more) general problem of statistical inference, where the unknown distribution on the pathways $\{p_c\}_{c=1}^C$ must be determined on the basis of partial information and represented by the conditions

\begin{equation}
\sum_c p_c=1\:\:\mbox{and}\:\:\sum_c p_cQ_c^{\alpha}=\langle Q^{\alpha}\rangle,\:\alpha=1\dots M,
\label{eq.II}
\end{equation}

\noindent where the second equation in \ref{eq.II} is nothing else than eq. \ref{eq.2}, rephrased in more general terms (with $Q_c^\alpha$ replacing $A_{\alpha c}$ and $\langle Q^\alpha\rangle$ replacing $r_\alpha$). Eq \ref{eq.3a} can thus be rewritten as

\begin{equation}
\mathcal{L}\equiv I(\mathbf{p},\mathbf{q},\gamma)-\theta_0\left[\sum_c p_c-1\right]-\sum_\alpha\theta_\alpha\left[\sum_c p_cQ^{\alpha}_c-\langle Q^\alpha\rangle\right]
\label{eq.3b}
\end{equation}

\noindent and the probability coefficients are obtained by solving the system 

\begin{equation}
\frac{\partial\mathcal{L}}{\partial p_c}=0,\:\forall\:c.
\label{eq.11}
\end{equation}

The resolution of the system \ref{eq.11} gives us the desired coefficients $\{p_c\}_{c=1}^C$ as functions of the Lagrangean multipliers, $p_c=p_c(\vec{\theta}),\:\forall\:c$. Once found, the parametric probability coefficients must be substituted back into $\mathcal{L}$, in order to obtain a quantity which is a function of the unknowns solely: $\mathcal{L}(\vec{\theta})$. The last step in the procedure is the optimization of the function $\mathcal{L}(\vec{\theta})$, by finding the values of the parameters $\vec{\theta}^*$ which satisfy the condition 

\begin{equation}
\frac{\partial\mathcal{L}}{\partial{\theta_i}}\bigg|_{\vec{\theta}^*}=0,\:\forall\:i.
\end{equation}

For expository purposes, we explicitly demonstrate the analytical derivation of the Shannon functional for univariate data sets. In this case, the probability coefficients given by eq. \ref{eq.11} have the expression

\begin{equation}
\frac{\partial\mathcal{L}}{\partial p_c}=0\Longrightarrow p_c=q_c\left(e^{-1+\theta_0+H_c}\right),\:\forall\:c
\end{equation}

\noindent having defined $H_c\equiv\sum_\alpha\theta_\alpha Q^\alpha_c$. Our probability coefficients can be thus rewritten as

\begin{equation}
p_c=\frac{q_ce^{\sum_\alpha\theta_\alpha A_{\alpha c}}}{\sum_c q_ce^{\sum_\alpha\theta_\alpha A_{\alpha c}}},\:\forall\:c.
\end{equation}

Substituting the analytical expression of $p_c$ back into $\mathcal{L}$ produces a quantity which is solely function of the vector of unknown parameters $\vec{\theta}$ and the function to optimize with respect to the vector $\vec{\theta}$ becomes

\begin{equation}
\mathcal{L}(\vec{\theta})=-\ln\left[\sum_c q_c\left(e^{\sum_\alpha\theta_\alpha A_{\alpha c}}\right)\right]+\sum_\alpha\theta_\alpha r_\alpha.
\end{equation}

\subsection{A second worked-out example concerning univariate data sets}

\begin{figure}[t!]
\centering
\includegraphics[width=0.9\textwidth]{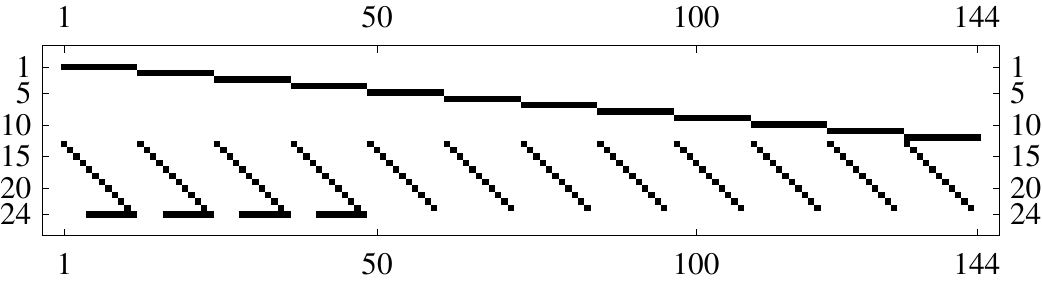}
\caption{Pictorial matrix representation of a local area network at the Information Networking Institute of Carnegie Mellon University (black squares represent ones, white squares represent zeros - see \cite{Airoldi}), composed by twelve subnetworks, communicating via two routers (one with four subnetworks, the second one with the remaining eight subnetworks - the routers are linked via a single connection). The network topology we consider yields 24 observed aggregate traffic volumes and 144 origin-destination traffic volumes to be estimated.}
\label{fig2b}
\end{figure}

For completeness, we discuss a second example of traffic networks. The data set was collected at the Information Networking Institute of Carnegie Mellon University (see \cite{Airoldi}) whose routing matrix is reported in fig. \ref{fig2b}. The network topology we consider yields 24 observed aggregate traffic volumes and 144 origin-destination traffic volumes, observed every five minutes (473 points in time). This second dataset is larger than the first, allowing us to test the scalability of our approach.

\begin{figure}[t!]
\centering
\includegraphics[width=0.49\textwidth]{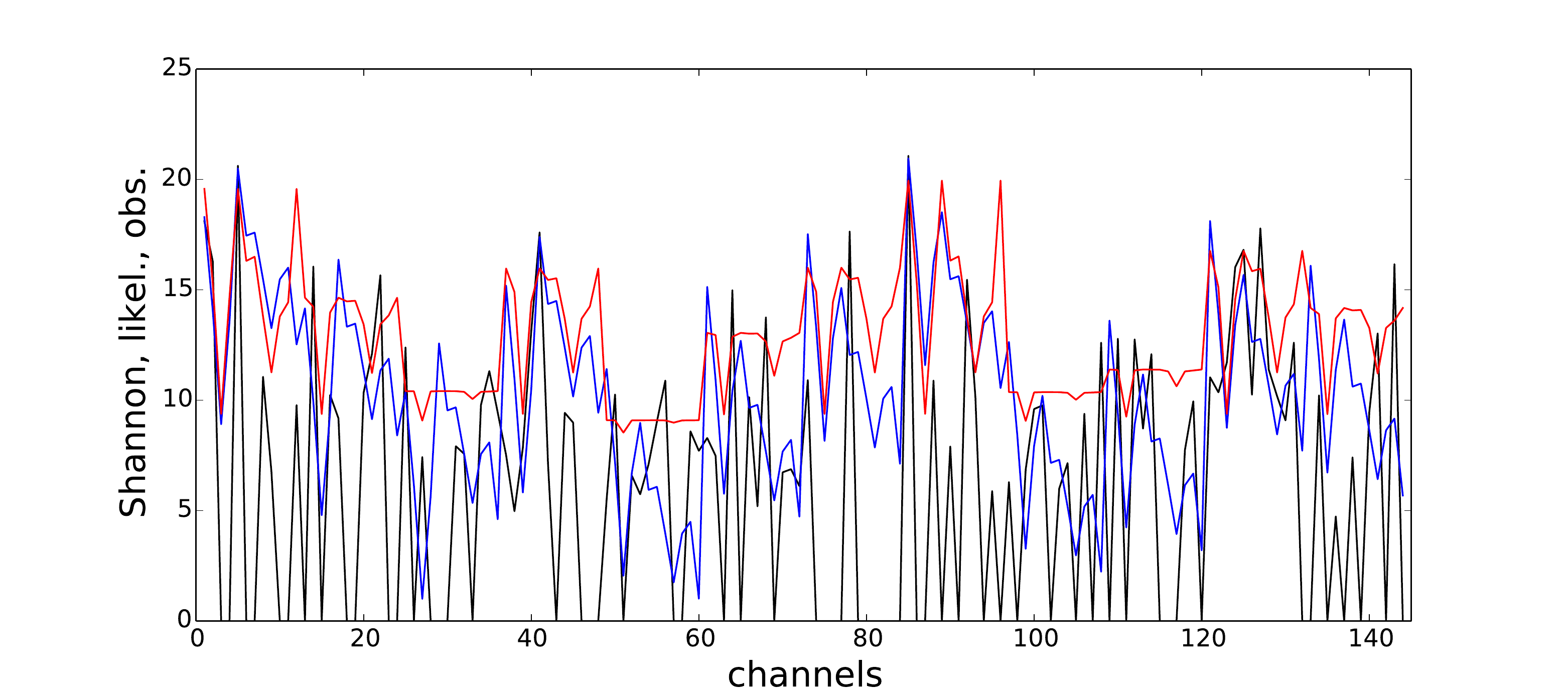}
\includegraphics[width=0.49\textwidth]{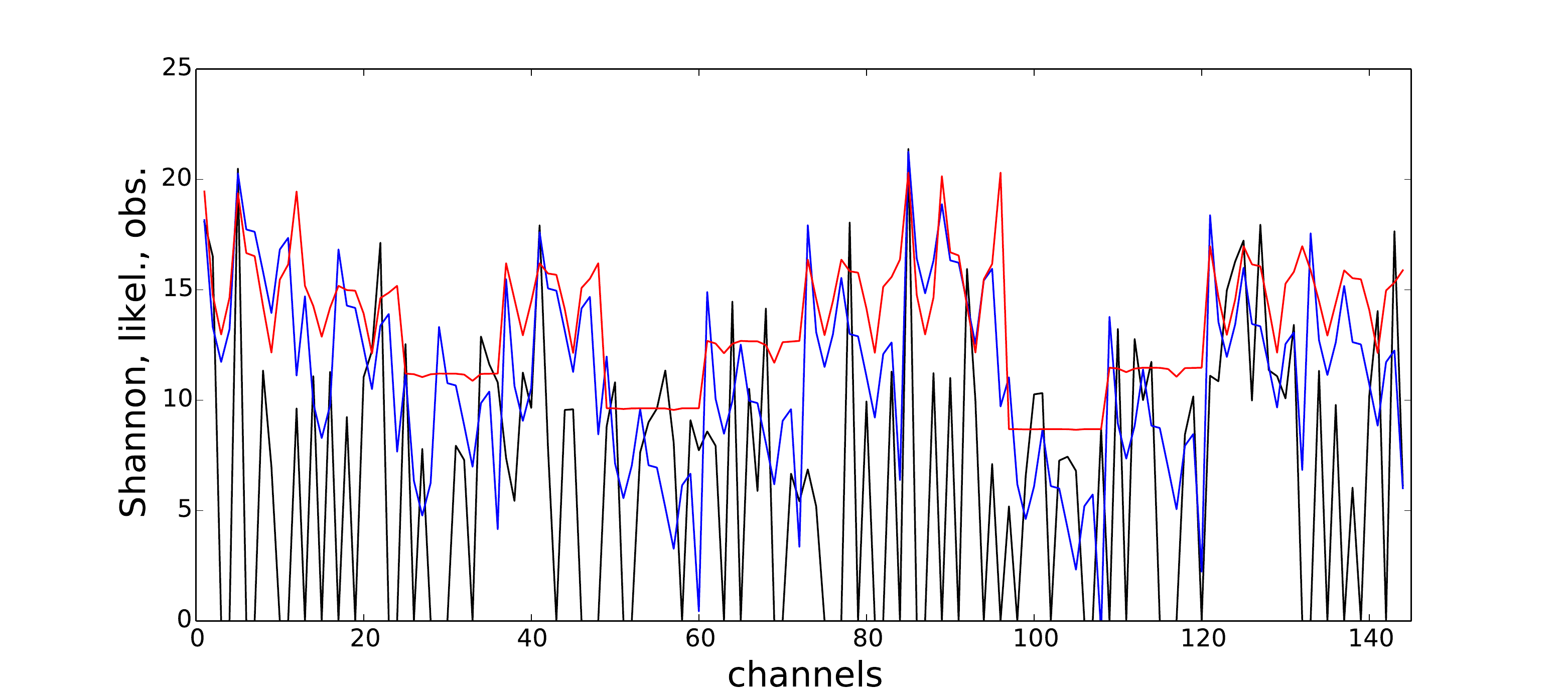}
\includegraphics[width=0.49\textwidth]{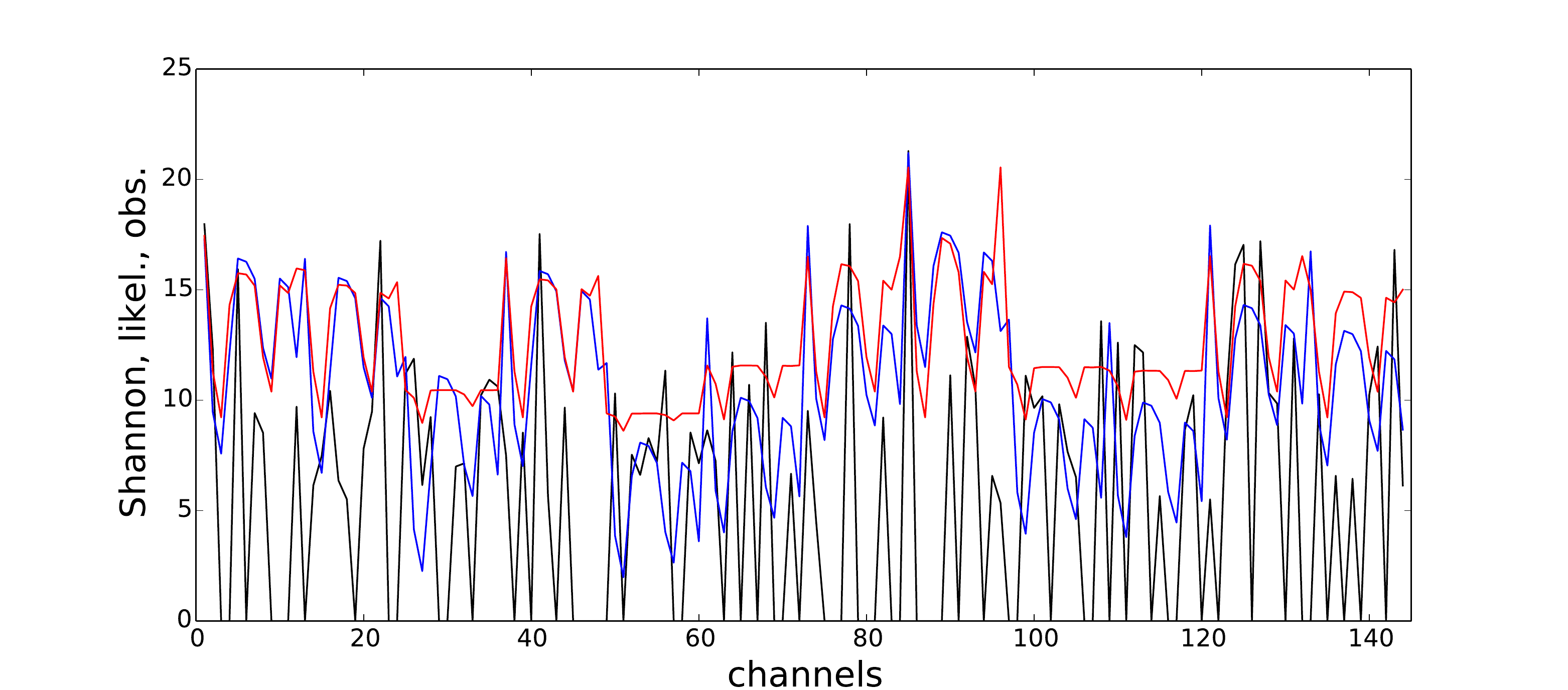}
\includegraphics[width=0.49\textwidth]{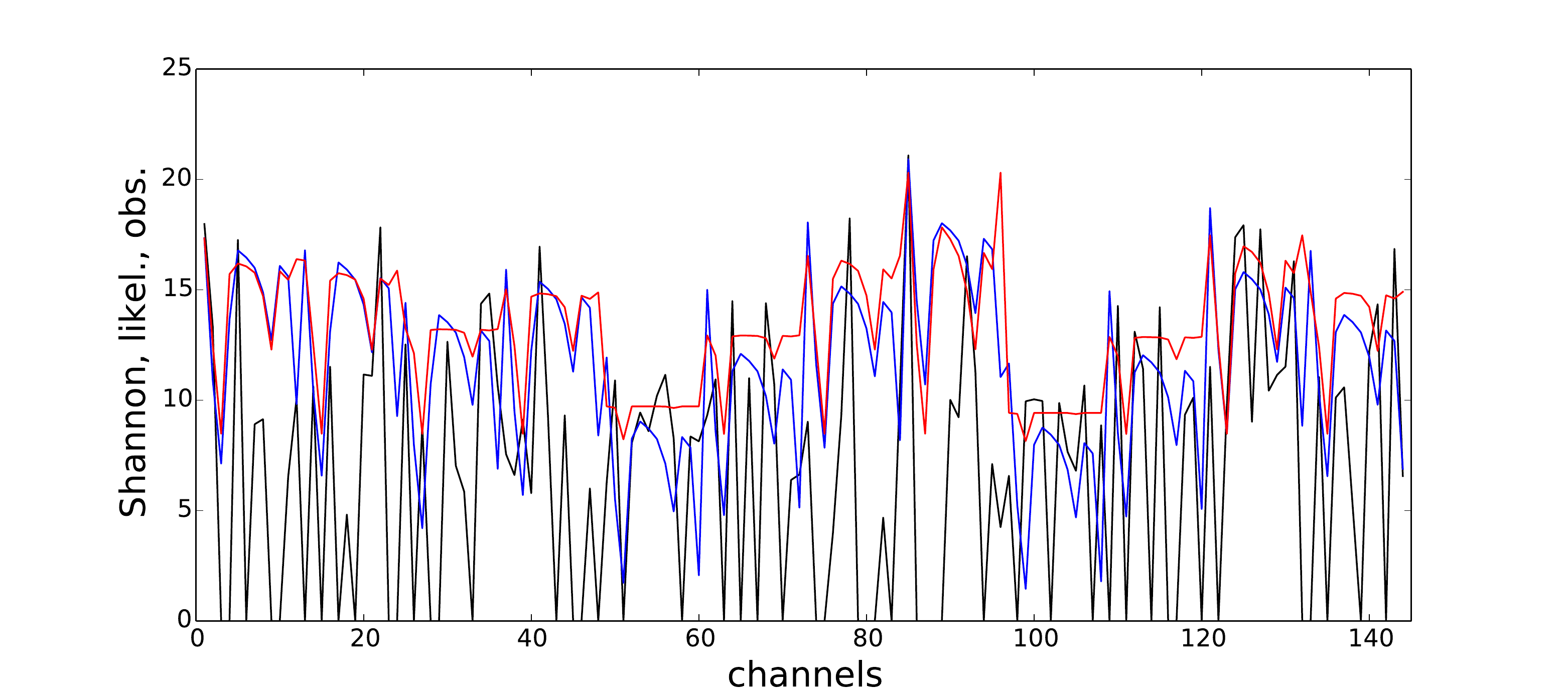}
\includegraphics[width=0.49\textwidth]{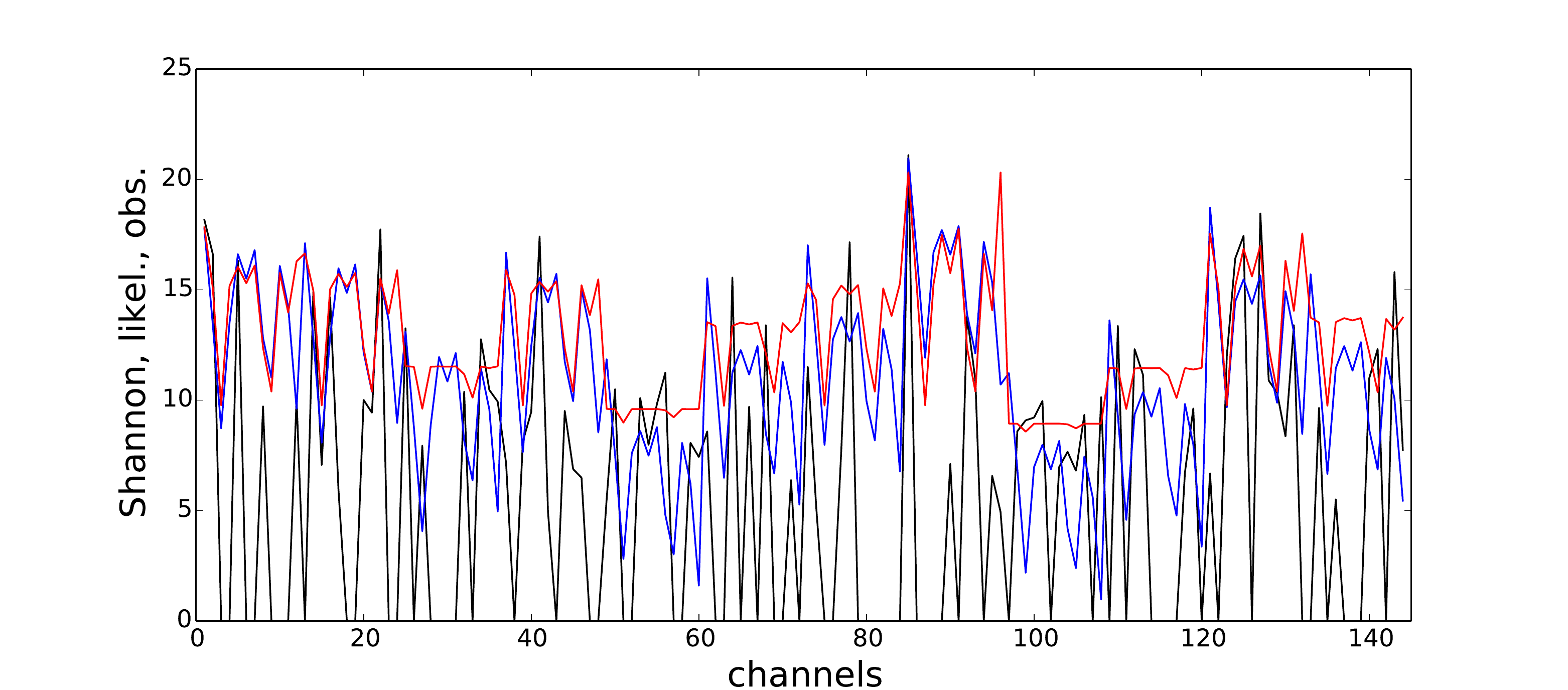}
\includegraphics[width=0.49\textwidth]{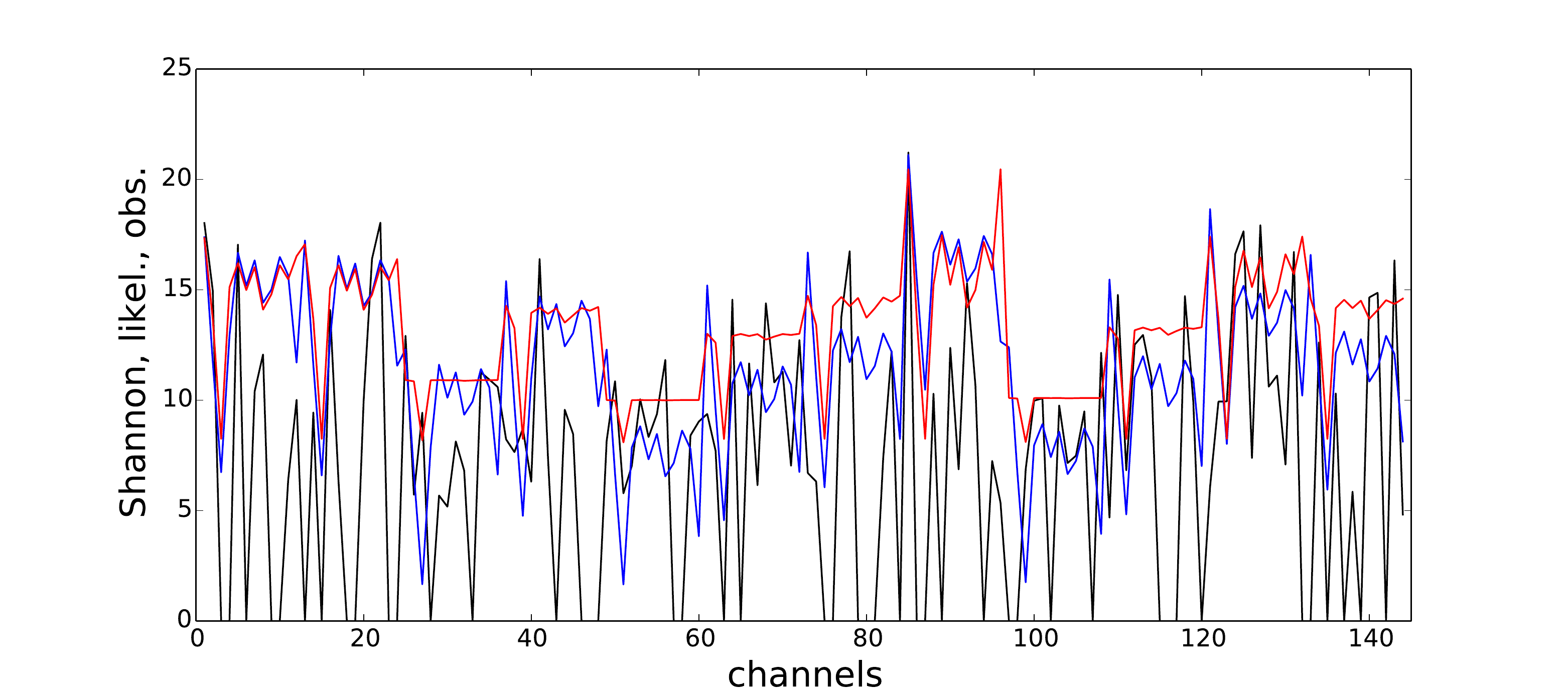}
\includegraphics[width=0.49\textwidth]{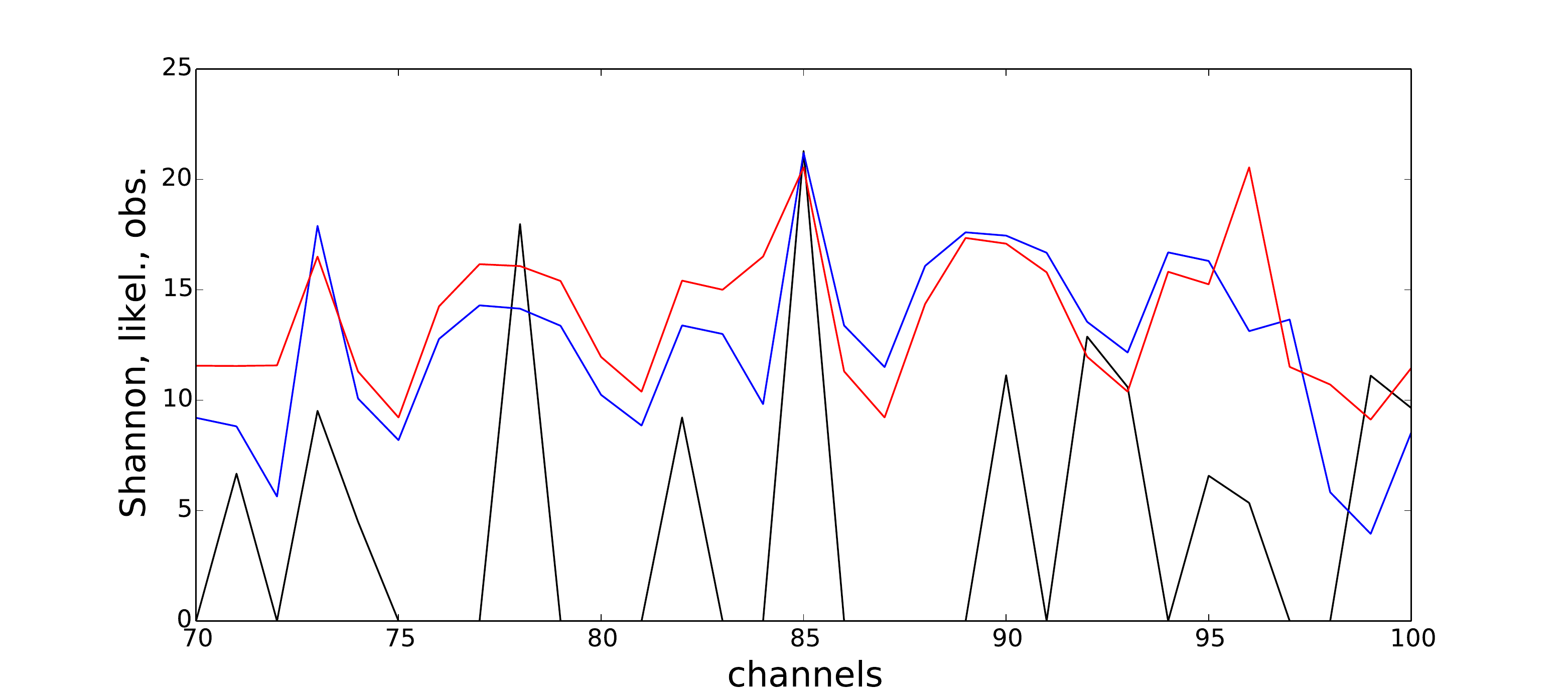}
\includegraphics[width=0.49\textwidth]{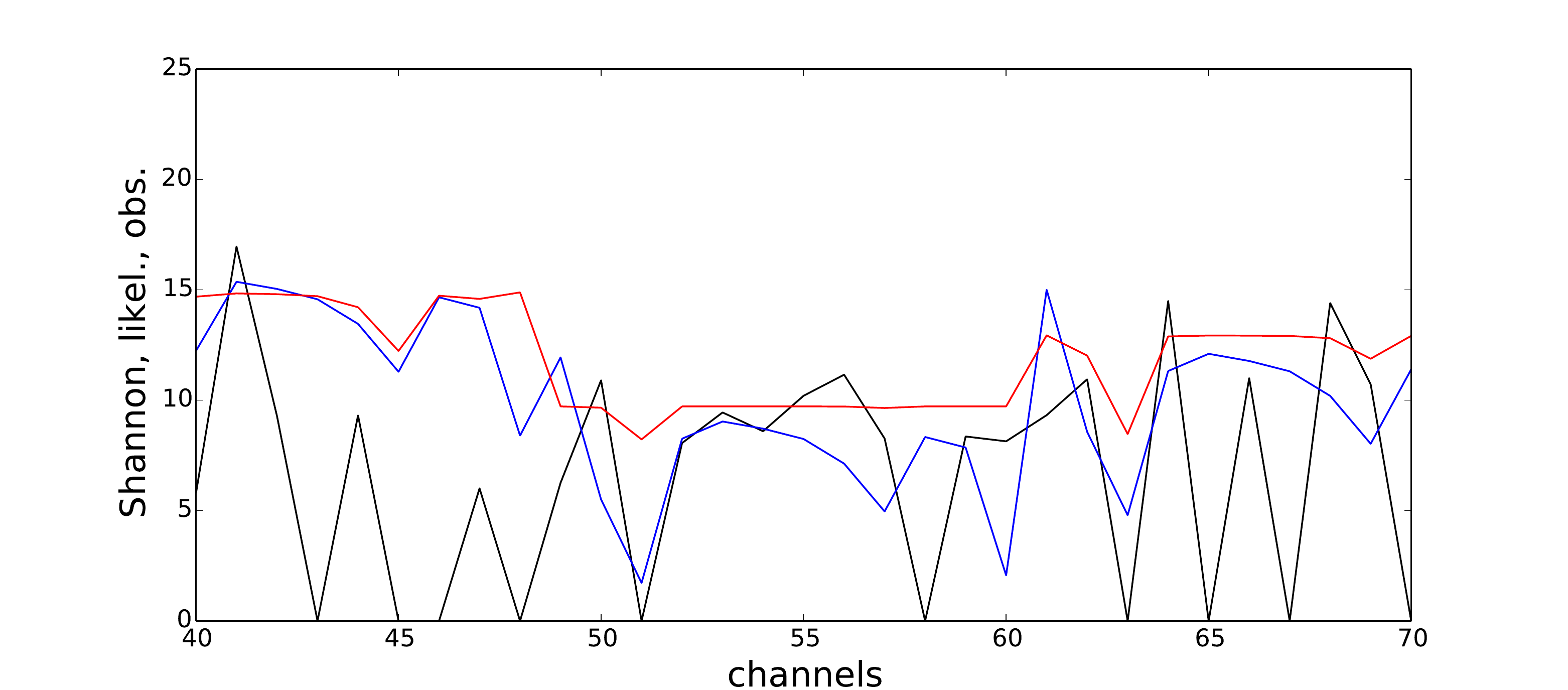}
\caption{Analysis of Carnegie University data corresponding to six chosen time points. The number of the channel is reported on the x-axis. Observed and estimated $\mathbf{x}$ are reported on the y-axis (logarithmic scale). Colors referes to: observed data (black trend), our estimation based on Shannon functional (blue trend), our estimation based on the likelihood functional (red trend). The lowest panels show a zoomed region of the ``channel plots'' corresponding to the 80th and 190th time points.}
\label{fig6}
\end{figure}

The analysis of Carnegie University data is illustrated in fig. \ref{fig6}. Again, our method captures the chosen temporal trends, impliying that our procedure is applicable to problems with higher dimensionality. However, the results concerning Carnegie University data present some differences with respect to the Bell Labs ones.

Since a visual inspection of fig. \ref{fig6} is not feasible, to quantify the agreement between our estimates and the observations we have calculated the correlation coefficient between the observed trends and the corresponding expected ones. The results for the Shannon functional are: $r=0.994$ for the upper left panel (1st time point), $r=0.991$ for the upper right panel (3rd time point), $r=0.996$ for the middle left panel (80th time point), $r=0.985$ for the middle right panel (190th time point), $r=0.989$ for the bottom left panel (330th time point) and $r=0.993$ for the bottom right panel (456th time point). The results for the likelihood functional are (in the same order): $r=0.581$ (1st time point), $r=0.595$ (3rd time point), $r=0.703$ (80th time point), $r=0.699$ (190th time point), $r=0.693$ (330th time point) and $r=0.701$ (456th time point).

Despite the rather high values of $r$, the strongly oscillatory character of the observed data set seems to have the effect of lowering the performance of our procedure: in fact, our estimations predict a ``smoother'' behavior than that of real data which, on the other hand, appear much more irregular (see lowest panels of fig. \ref{fig6}). As for the Bell Labs data set, the net result is that high values of traffic data are well estimated while the lower ones (included the zero ones) are generally overestimated. 

Quite surprisingly, even the differences characterizing the performances of the two functionals are larger than for the Bell Labs data set: this time the best result (witnessed by the higher correlation coefficients for all the time points) is obtained by the Shannon functional which seems to better follow the irregular observed trends: the predictions obtained by the likelihood functional, in fact, show flat regions which in turn have the effect of lowering the numerical correlation value.

\subsection{Bivariate data sets}

For bivariate problems, the CR family of functionals becomes

\begin{equation}
I(\mathbf{p}, \mathbf{q}, \gamma)=\frac{1}{\gamma(\gamma+1)}\sum_j\sum_kp_{jk}\left[\left(\frac{p_{jk}}{q_{jk}}\right)^\gamma-1\right]
\label{shaeq.2}
\end{equation}

\noindent $j$ and $k$ respectively indicating the row and column index of the probability matrix $\mathbf{P}$ to be estimated and of the prior, bivariate one $\mathbf{Q}$. The constraints are now represented by the conditions

\begin{equation}
\sum_{k} p_{jk}=1,\:\forall\:j\:\:\mbox{and}\:\:\sum_{j} x_j'p_{jk}=y_k',\:\forall\:k.
\end{equation}

For bivariate problems, the number of multipliers rises, since the required number of normalization conditions equals the number of matrix rows. Thus, in order to correctly implement our approach, two vectors $\vec{\alpha}$ and $\vec{\beta}$ must be considered. Constraining equation \ref{shaeq.2} for bivariate data sets (and again for Shannon entropy) leads to

\begin{equation}
I\left(\mathbf{p}, \frac{1}{C}, 0\right)=\sum_j\sum_kp_{jk}\ln p_{jk}+\ln C-\sum_j\beta_j\left(\sum_kp_{jk}-1\right)-\sum_k\alpha_k\left(\sum_jp_{jk}x_j'-y_k'\right)
\end{equation}

\noindent and maximizing it with respect to $p_{jk}$ implies that the functional form of our coefficients is

\begin{equation}
p_{jk}=\frac{e^{\alpha_k x_j'}}{\sum_k e^{\alpha_k x_j'}},\:\forall\:j,k;
\end{equation}

\noindent by substituing back into $\mathcal{L}$ we get

\begin{equation}
\mathcal{L}(\vec{\alpha})=-\sum_j\left(\ln\left[\sum_k e^{\alpha_k x_j'}\right]+\sum_k\alpha_kx_j'\right).
\end{equation}

Similar results are obtained for the other functionals.

\subsection{A second worked-out example concerning bivariate data sets}

The second bivariate data set we discuss comes from an application in political science and concerns voter behavior and candidate choice (as reported in table \ref{table2} - see \cite{Cho2006}). The result of the application of our method to the elections percentages is shown in table \ref{table4}. 

Since privacy issues prevent the percentage of people voting for a given candidate from being available, the second bivariate data set we analyzed provides only aggregate data about the elections results: the single matrix entries are thus missing. Nonetheless, our method provides a prediction of the unknown entries, by adopting the same procedure used for the ``eggs and bacon'' problem. As can be seen from table \ref{table4}, Shannon functional and the likelihood functional give compatible estimates of the voting percentages: this similarity is effectively summed up by the ``global'' Pearson correlation coefficient between the Shannon expected matrix and the likelihood expected matrix (both treated as an unique vector of numbers), equal to 0.988716. It should be noted, however, that significative differences can be observed for the percentages referring to the independent candidates. Nonetheless, when interpreted in the light of the previous results, these differences carry an important information, signalling that independent candidates true percentages are, probably, not only the lowest ones, but even compatible with zero.

\begin{table}[t!]
\centering
\caption{Precint-level data of Louisiana's 5th CD elections (see \cite{Cho2014}.}
\begin{tabular}{c|ccccc|c}
\noalign{\smallskip}
$$ & $\mbox{Rep.}$ & $\mbox{Dem.}$ & $\mbox{Ind.1}$ & $\mbox{Ind.2}$ & $\mbox{Abst.}$ & $\mbox{Total}$\\
\hline
$\mbox{White}$ & $-$ & $-$ & $-$ & $-$ & $-$ & $1158$\\

$\mbox{Black}$ & $-$ & $-$ & $-$ & $-$ & $-$ & $222$\\

$\mbox{Other}$ & $-$ & $-$ & $-$ & $-$ & $-$ & $31$\\
\hline
$\mbox{Total}$ & $963$ & $207$ & $28$ & $17$ & $196$ & $1411$
\end{tabular}
\label{table2}
\end{table}

\begin{table*}[h!]
\centering
\caption{Estimated precint-level percentages of Louisiana's 5th CD elections (see \cite{Cho2014}).}
\begin{tabular}{|c|ccccc|c|}
\hline
$$ & $$ & $$ & $\mbox{Shannon functional}$ & $$ & $$ & $$\\
\hline
$$ & $\mbox{Rep.}$ & $\mbox{Dem.}$ & $\mbox{Ind.1}$ & $\mbox{Ind.2}$ & $\mbox{Abst.}$ & $\mbox{Total}$\\
\hline
$\mbox{White}$ & $877.555$ & $144.824$ & $0.968424$ & $0.0422665$ & $134.611$ & $1158$\\

$\mbox{Black}$ & $78.5616$ & $55.6173$ & $21.2953$ & $11.6828$ & $54.843$ & $222$\\

$\mbox{Other}$ & $6.88327$ & $6.55916$ & $5.73627$ & $5.27497$ & $6.54634$ & $31$\\
\hline
\hline
$$ & $$ & $$ & $\mbox{Likelihood functional}$ & $$ & $$ & $$\\
\hline
$$ & $\mbox{Rep.}$ & $\mbox{Dem.}$ & $\mbox{Ind.1}$ & $\mbox{Ind.2}$ & $\mbox{Abst.}$ & $\mbox{Total}$\\
\hline
$\mbox{White}$ & $865.704$ & $141.143$ & $12.2831$ & $6.89041$ & $131.831$ & $1158$\\

$\mbox{Black}$ & $89.4101$ & $58.4307$ & $10.9359$ & $6.44502$ & $56.7707$ & $222$\\

$\mbox{Other}$ & $7.7549$ & $7.41397$ & $4.77993$ & $3.66401$ & $7.38656$ & $31$\\
\hline
$\mbox{Total}$ & $963$ & $207$ & $28$ & $17$ & $196$ & $1411$\\
\hline
\end{tabular}
\label{table4}
\end{table*}

\section*{Acknowledgements}
TS acknowledges support from the Italian PNR project CRISIS-Lab. 

ESG acknowledges support from the European Commission Marie-Curie ITN program (FP7-320 PEOPLE-2011-ITN) through the LINC project (no. 289447). 

DG acknowledges support from the Dutch Econophysics Foundation (Stichting Econophysics, Leiden, the Netherlands). This work was also supported by the EU project MULTIPLEX (contract 317532) and the Netherlands Organization for Scientific Research (NWO/OCW).

\end{document}